\documentstyle[prl,aps,twocolumn,psfig,floats]{revtex}
\begin{document}

\title{Statistical signatures of critical behavior in small systems}

\author{
J. B. Elliott$^1$, 
S. Albergo$^2$, 
F. Bieser$^6$, 
F. P. Brady$^3$, 
Z. Caccia$^2$,\\
D. A. Cebra$^3$, 
A. D. Chacon$^7$, 
J. L. Chance$^3$, 
Y. Choi$^{1}$, 
S. Costa$^2$,\\
M. L. Gilkes$^{1}$, 
J. A. Hauger$^1$, 
A. S. Hirsch$^1$, 
E. L. Hjort$^1$, 
A. Insolia$^2$,\\
M. Justice$^5$, 
D. Keane$^5$, 
J. C. Kintner$^3$, 
V. Lindenstruth$^{4}$, 
M. A. Lisa$^6$,\\
H. S. Matis$^6$, 
M. McMahan$^6$, 
C. McParland$^6$, 
W. F. J. M\"{u}ller$^4$, 
D. L. Olson$^6$,\\
M. D. Partlan$^{3}$, 
N. T. Porile$^1$, 
R. Potenza$^2$,
G. Rai$^6$, 
J. Rasmussen$^6$,\\
H. G. Ritter$^6$, 
J. Romanski$^{2}$, 
J. L. Romero$^3$, 
G. V. Russo$^2$,
H. Sann$^4$,\\
R. P. Scharenberg$^1$, 
A. Scott$^5$, 
Y. Shao$^{5}$, 
B. K. Srivastava$^1$, 
T. J. M. Symons$^6$,\\
M. Tincknell$^1$, 
C. Tuv\'{e}$^2$, 
S. Wang$^5$, 
P. G. Warren$^1$, 
H. H. Wieman$^6$,\\
T. Wienold$^6$, and 
K. Wolf$^7$\\
(EOS Collaboration)
}

\address{
$^1$Purdue University, West Lafayette, IN 47907\\ 
$^2$Universit\'{a} di Catania and Istituto Nazionale di Fisica Nucleare-Sezione di Catania, 95129 Catania, Italy\\ 
$^3$University of California, Davis, CA 95616\\ 
$^4$GSI, D-64220 Darmstadt, Germany\\ 
$^5$Kent State University, Kent, OH 44242\\ 
$^6$Nuclear Science Division, Lawrence Berkeley National Laboratory, Berkeley, CA 94720\\ 
$^7$Texas A\&M University, College Station, TX  77843
}

\date{\today}

\maketitle

\begin{abstract}
The cluster distributions of different systems are examined to search for signatures of a continuous phase transition.\ \ In a
system known to possess such a phase  transition, both sensitive and insensitive signatures are present; while in systems known
not to possess such a phase transition, only insensitive signatures are present.\ \ It is shown that nuclear multifragmentation
results in cluster distributions belonging to the former category, suggesting that the fragments are the result of a continuous 
phase transition.\\
PACS number(s): 25.70 Pq, 64.60.Ak, 24.60.Ky, 05.70.Jk
\end{abstract}

\narrowtext

\section{Introduction}

Beginning in the 1970's significant advances in the understanding of  nuclear multifragmentation were made possible with the advent
of high statistics inclusive experiments.\ \ Typically, only one intermediate mass fragment ( $3 \le Z_f \le 30$ ) was detected per
event.\ \ From these inclusive studies came the first evidence that intermediate mass fragments (IMFs) were associated with a
simultaneous multi-body breakup of a system which had undergone expansion.\ \ A study of the fragment mass yield distribution
obtained in an inclusive gas jet experiment conducted at Fermilab contained the first indication that nuclear multifragmentation
might be related to critical phenomena normally observed in macroscopic systems \cite{finn}.\ \  The Purdue Group was the first to
make the suggestion that the observed power law in the fragment yield distribution might result from a system whose excitation
energy was comparable to its total binding energy \cite{hirsch}.\ \ The exponent of the power law was $2 {\le} {\tau} {\le} 3$,
within the range expected for a system near its critical point.\ \ The presence of the power law and the value of the exponent,
coupled with the strong similarity of the nuclear and van der Waals potentials, led the Purdue group to suggest that
multifragmentation of nuclei might be analogous to a fluid undergoing a continuous phase transition from a liquid to a gas.\ \
Furthermore, the Fisher Droplet Model (FMD) \cite{fisher}-\cite{stauffer_kiang_2}, used to describe condensation in a fluid system
near its critical point, after modification for nuclear physics effects, was capable of describing the isotopic yields of 50
fragments with one set of parameters \cite{hirsch}, \cite{minich}.\ \ The temperature of the system was determined to be about 5
MeV \cite{hirsch}, a reasonable value considering that the average binding energy per nucleon in a nucleus is approximately 8 MeV.\ \ 
The success of this approach reinforced the notion that  multifragmentation was both a thermal process and that it was related to 
critical phenomena.

With the advent of exclusive experiments capable of detecting all of the charged reaction products, the possibility of studying
multifragmentation on an event-by-event basis became a reality.\ \ High statistics exclusive experiments in which the fragmenting
system is  characterized according to its nucleon number and excitation energy permit both the correlation of dynamical and
statistical information and the study of fluctuations in experimental observables.\ \ Fluctuations are central to all critical
phenomena, and indeed, such fluctuations are apparent in exclusive multifragmentation data.\ \ In this paper, the focus will be on
the statistical  signals of multifragmentation data observed in the EOS experiment \cite{gilkes_gamma}-\cite{elliott_scaling}.\ \
Comparisons will be made with two other systems, one of which exhibits critical behavior and one of which does not. 

Much of the pioneering work in understanding the statistical aspects of multifragmentation has been performed by Campi 
\cite{campi_1}-\cite{campi_8} and Mekjian \cite{mekjian_1}-\cite{mekjian_7}.\ \ Both efforts have compared multifragmentation data
to model systems in order to gain some insight into the nuclear breakup process.\ \ In this paper, many of the ideas suggested by
these authors are followed and applied to both the EOS data and the model systems in order to demonstrate which of the many
suggested signals are useful for the identification of critical behavior.\ \  A major goal of this paper is to present a 
comprehensive review of several methods proposed for detecting signals of critical phenomena in multifragmentation.

It is tempting to compare the experimental data to dynamical models that attempt to describe nuclear multifragmentation.\ \
However, the task of modeling multifragmentation from the initial collision phase of the reaction to freeze-out has proven to be a
daunting task.\ \ Models that adequately describe the initial stage of the reaction \cite{yariv_1}-\cite{peilert} do not
satisfactorily describe the fragment formation stage, in either statistical or dynamical aspects.\ \  Likewise, the most successful
models in describing the statistical properties of nuclear multifragmentation \cite{bondorf_1}-\cite{gross_2}, assume thermodynamic
equilibrium, yet fail to adequately match the dynamical features of the data.

Molecular dynamical approaches, which have enjoyed considerable success in describing critical behavior in classical systems
\cite{schlagel}-\cite{pratt}, have not been conclusive in describing nuclear multifragmentation and at times have yielded
contradictory results \cite{pratt}, \cite{latora}.\ \ Later studies suggested flaws in the application of molecular dynamical
models to nuclear multifragmentation, therefore calling into question the conclusions drawn from the earlier studies
\cite{donangelo}.

The most striking of the early theoretical efforts came from Campi's analysis of a few hundred completely reconstructed emulsion
multifragmentation events \cite{waddington_friar} and the comparison of these data to clusters generated from a percolation
calculation \cite{campi_1}, \cite{campi_2}.\ \ In this series of papers it was shown that the fragment distributions from
multifragmentation bore a striking similarity to the cluster distributions from percolation lattices.\ \ This analysis provided
strong evidence that multifragmentation was a statistical process which appeared to be related to critical phenomena.\ \ In this
analysis another estimate of the exponent ${\tau}$ was made which agreed with the first measurements from the Purdue Group and
several later analyses of various fragment distributions.

In the early 1990's the ALADIN Group from GSI performed several multifragmentation experiments \cite{gsi_1}-\cite{gsi_3}.\ \ Of
particular importance was the ``rise and fall'' of multifragmentation.\ \ In one analysis the ALADIN group plotted the ``rise
and fall'' curve of the production of IMFs versus an observable related to the excitation energy of the reaction for several
multifragmenting systems.\ \ With the appropriate scaling the data collapsed to a single curve suggesting that the
multifragmenting systems retained no memory of the reaction entrance channel.\ \ This is expected for an equilibrated system.

The results of some statistical analyses of multifragmentation data could be interpreted to suggest that multifragmentation is a
sequential decay \cite{moretto} in contrast to the phase transition picture.\ \ The same sort of statistical analysis has also been
applied to explicitly simultaneous models \cite{donangelo} and produced results that were similar to those of multifragmentation
data.\ \ Thus those signals could be interpreted as evidence for either sequential or simultaneous multifragmentation \cite{moretto}.

This last effort puts into focus the main question in this work: what type of analysis of the statistical aspects of a cluster
distribution can provide the most insight into the nature of the mechanism which created the clusters?\ \ Specifically, can those 
systems which contain critical behavior be distinguished from those which do not?\ \ It will be argued that this question has two 
answers.\ \ Analysis of the {\it insensitive} features of the cluster distribution cannot make the above mentioned distinction 
\cite{phair}.\ \ However, an analysis of the {\it sensitive} features of the cluster distribution will be shown to provide deeper 
insight into the cluster production mechanism.\ \ This type of analysis has been previously reported for clusters resulting from 
nuclear multifragmentation \cite{gilkes_gamma}-\cite{elliott_scaling}.\ \ Note that the more generic term {\it cluster} will be used 
to refer to any composite of constituents, whether these be molecules of a fluid, nuclear fragments or percolation clusters.

The method employed to address the above question is as follows.\ \ The same analysis is performed on the cluster distributions
produced by three different systems.\ \ In one case, clusters are generated by randomly partitioning an integer.\ \ Such
one-dimensional partitioning does not posses critical behavior indicative of a continuous phase transition.\ \ In the second case,
three-dimensional bond building percolation is used to produce clusters.\ \ Percolation is well-known mathematical construct that 
possesses a continuous phase transition, {\it i.e.} a critical point.\ \ Finally, the cluster distributions resulting from the 
multifragmentation of gold nuclei are analyzed.\ \ Although it is not known, a priori, whether the nuclear multifragmentation bears 
any relation to critical phenomena, it will be seen that the analysis presented in this work yields suggestive results.

This paper is organized as follows.\ \ In section II a description of each system is presented.\ \ In section III the Fisher
Droplet Model is reviewed.\ \ In section IV-A the insensitive signatures of the cluster distributions for all systems are examined.\
\ In section IV-B the sensitive signatures are examined.\ \ Sections V and VI present possible corrections to the analysis of the
multifragmentation data.\ \ Finally, Section VII discusses the conclusions reached upon the completion of the analyses in sections
IV and V. Throughout this paper the term {\it continuous phase transition} will be used instead of {\it second order phase
transition}, the latter from the outdated Ehrenfest theory of phase transitions \cite{theory_of_critical_phenomena}.

\section{Description of systems under study}

\subsection{1.0 A GeV Au $+$ C multifragmentation}

Approximately 40,000 fully reconstructed events ($76 \le Z_{observed} \le 82$) were collected with the EOS experimental apparatus
discussed in ref. \cite{gilkes_gamma}.\ \ In the collision of the projectile gold nucleus ($197$, $79$) and the target carbon nucleus, 
so-called prompt nucleons are knocked out of the gold nucleus by quasi-elastic and inelastic collisions between projectile and target 
nucleons \cite{hauger_prl}.\ \ Immdeiately following the collision, he gold projectile remnant is in an excited state with fewer than 
197 nucleons.\ \ The excited remnant cools and expands, evolving to the neighborhood of the critical point in the temperature-density 
plane \cite{hauger_prc}, where clusters condense from a high temperature low density vapor of nucleons.\ \  The charge and mass 
of the projectile remnant, $Z_{0}$ and $A_{0}$, were determined for each event by subtracting the charge and mass of the prompt 
particles from the charge and mass of the gold nucleus \cite{hauger_prc}.\ \ Prompt particles have $Z_f = 0$, $1$ and $2$ and are 
removed from the cluster distributions analyzed in this work.\ \ Only clusters created from the excited gold projectile are considered 
in the ensuing analysis.\ \ For events with the lowest charged particle multiplicities, $m$, the remnant had $Z_{0} \sim 76$, 
$A_{0} \sim 194$ and $E^{\ast}/A_{0} \sim 2$ MeV$/$nucleon, while for events with the highest multiplicities the remnant had 
$Z_{0} \sim 39$, $A_{0} \sim 92$ and $E^{\ast}/A_{0} \sim 16$ MeV$/$nucleon \cite{hauger_prc}.

Clusters of a given charge, $Z_f$, were counted on an event by event basis to determine the cluster charge distribution,
$N_{Z_{f}}$.\ \ In this analysis, although the mass number of the clusters is of interest, a cluster's charge will be used as an
index.\ \ Mass numbers for clusters of charge one and two were measured in the EOS time projection chamber.\ \ Clusters with $Z_f
\ge 3$ were assigned a mass number, $A_f$, by multiplying the cluster charge by the mass to charge ratio of the excited gold
projectile remnant; for low $m$ events $A_{0}/Z_{0} \sim 2.55$ and for high $m$ events $A_{0}/Z_{0} \sim 2.36$.\ \ This procedure
provided an estimate of a cluster's mass number prior to any secondary decay effects.\ \ It is assumed that on average 
$N_{A_{f}} = N_{Z_{f}}$.\ \ Finally, it is the normalized cluster distribution, $n_{A_{f}} = N_{A_{f}} / A_{0}(m)$, that is used 
in the analysis presented in this paper.

\subsection{Percolation}

Bond building percolation calculations were performed on three dimensional simple cubic lattices of 216 sites.\ \ Cluster
distributions for $100,000$ lattice realizations were generated in the standard fashion by forming bonds between sites.\ \ Bonds
were either active (on) or inactive (off) according to the following algorithm.

The control parameter ({\it e.g.} temperature in thermodynamic systems) for percolation is the lattice probability, $p_l$.\ \ A
single value of $p_l$ was chosen for the entire lattice.\ \ All probabilities were between 0 and 1.\ \ Next, a bond probability,
${p_{b_{i}}}$, was randomly chosen from a uniform distribution on (0,1) for the $i^{th}$ bond.\ \ If ${p_{b_{i}}}$ was less than
$p_l$, then the $i^{th}$ bond was active and two sites were joined into a cluster.\ \ This process was performed for each bond in
the lattice.

At low values of $p_l$, few bonds were formed resulting in a high multiplicity, $m$, of small clusters, a distribution analogous 
to the gaseous phase of a fluid.\ \ At high values of $p_l$, many bonds were formed resulting in a low multiplicity of mostly large 
clusters, analogous to the liquid phase of a fluid.\ \ In an infinite lattice the phase transition occurs at a unique value of the 
lattice probability, $p_c$, when the probability of forming a percolating cluster changes from zero to unity.

To examine the behavior of the average cluster distribution, the number of clusters of size $A_f$ per lattice site was calculated
by histogramming the $100,000$ lattice realizations into 100 bins from $0$ to $1$.\ \ The use of $m$ as a control parameter and the
ensuing effects on signatures of continuous phase transition were investigated by calculating the average number of clusters of
size $A_f$ with the $100,000$ lattice realizations histogrammed in units of $m$.\ \ 

\subsection{Random partitions}

Random partitions were generated from 79 total system constituents, chosen to approximate the number of charges in the gold
multifragmentation system.\ \ The algorithm is as follows.\ \ First a random choice of $m$ was made from a uniform distribution on 
(1,79).\ \ Next the maximum size of a cluster, $A_{max}^{1}$, for an event with $m$ was determined; this depended on the constraints 
of the system size, $A_0 = 79$ and the choice of $m$.\ \ The size of the first cluster, $A_1$, was then randomly chosen from a uniform 
distribution on ${(1,A_{max}^{1})}$.\ \ There were then $m-1$ clusters to be generated from $79-A_1$ constituents.\ \ The maximum size 
of a cluster for an $m-1$ event from a $79-1$ constituent system was determined: $A_{max}^{2}$.\ \ The size of the second cluster, 
$A_2$, was then randomly chosen form a uniform distribution on $(1,A_{max}^{2})$.\ \ This process was repeated until all constituents 
belonged to a cluster.\ \ $100,000$ partitions were generated in this manner.\ \ This particular weighting results in a power law 
cluster distribution \cite{mekjian_4}.

\section{Review of the Fisher Droplet Model}

The focus of most studies of phase transitions is on standard thermodynamical variables such as a system's temperature, density,
compressibility, etc.\ \ These quantities are difficult or impossible to measure directly in present nuclear multifragmentation
experiments.\ \ Thus a theory which addresses quantities accessible to MF experiments is needed.\ \ To that end Fisher's 
gas-to-liquid phase transition model, based on Mayer's condensation theory, is followed \cite{fisher}, \cite{stauffer_kiang},
\cite{stauffer_aharony}.

Fisher begins his model, called the Fisher Droplet Model (FDM) hereafter, by writing the free energy for the formation of clusters
of size $A_f$ as:
     \begin{eqnarray}
     \label{free_energy}
     {\Delta} G_{A_{f}} & = -  & k_{b} T A_{f} \ln(g({\mu},T)) - \nonumber \\
                        &      & k_{b} T \ln(f(A_{f},T)) + k_{b} T {\tau} \ln(A_{f}) + \ldots 
     \end{eqnarray}
Where $k_b$ is the Boltzmann constant and the $g$-term is the bulk formation energy, or volume term and:
     \begin{equation}
     g({\mu},T) = exp[({\mu}-{\mu}_{coex})/k_{b} T],
     \label{volume_term}
     \end{equation}
where $\mu$ is the chemical potential and ${\mu}_{coex}$ is the chemical potential along the coexistence curve.\ \

The $f$-term is related to the surface free energy of cluster formation.\ \ It's a form given by Fisher is:
     \begin{equation}
     f(A_{f},T) = exp[a_{0} {\omega} A_{f}^{\sigma} {\epsilon} T_{c} / 
     k_{b} T], 
     \label{surface_term_fisher}
     \end{equation}
where $\sigma$ is a critical exponent and is related to the ratio of the dimensionality of the surface to the dimensionality of the
volume, $a_0$ is a constant of proportionality relating the average surface area of a droplet to its number of constituents and
$\omega$ is the surface entropy density; $\epsilon$ is a measure of the distance from the critical point.\ \  For usual
thermodynamic systems ${\epsilon} = ( T_{c} - T ) / T_{c}$, in the percolation treatment ${\epsilon} = ( p_{l} - p_{c} ) / p_{c}$
and for multifragmentation ${\epsilon} = ( m_{c} - m ) / m_{c}$ will be used.\ \ All formulations of $\epsilon$ are such that
$\epsilon > 0$ ($\epsilon < 0$) corresponds to the {\it liquid} ({\it gas}) region.\ \ This form of the surface free energy is
applicable on only one side of the critical point, the single phase side.\ \ A more general form suggested by efforts from
percolation theory \cite{stauffer_rep}-\cite{leath_2} that can be applied on both sides of the critical point and leads to a power 
law which describes the behavior of the order parameter is:
     \begin{equation}
     f(z) = A exp[ -(z-B)^{2} / C ] ,
     \label{surface_term_leath}
     \end{equation}
where the scaling variable, $z$, is
     \begin{equation}
     z = A_{f}^{\sigma} {\epsilon}.
     \label{scaling_variable}
     \end{equation}
The physical interpretation of the parameters $A$, $B$ and $C$ is an open question.

Finally $\tau$ is another critical exponent depending principally on the dimensionality of the system and has its origins in
considerations of a three dimensional random walk of a surface closing in on itself.\ \ For three dimensions $2 \le {\tau} \le 3$
\cite{fisher_priv_com}.\ \ In eq. (\ref{free_energy}), $q_0$ is a normalization constant which will be shown to depend solely on
the value of $\tau$ \cite{nakanishi}.

From the free energy of cluster formation the average cluster distribution normalized to the size of the system is:
     \begin{equation}
     n_{A_{f}}({\epsilon}) = \exp (-{\Delta} G_{A_{f}} / k_{b} T ) = q_{0} A_{f}^{-{\tau}} f(z) {g({\mu},T)}^{A_{f}}. 
     \label{cluster_distribution}
     \end{equation}
At the critical point, ${\epsilon} = 0$, both $f$ and $g$ are unity and the cluster distribution is given by a pure power law:
     \begin{equation}
     n_{A_{f}}({\epsilon}) = q_{0} A_{f}^{-{\tau}}.
     \label{power_law}
     \end{equation}
If the first moment of the normalized cluster distribution is considered at the critical point then
\cite{nakanishi}:
     \begin{equation}
     M_{1}({\epsilon} = 0) = {\sum}_{A_f} n_{A_{f}}({\epsilon}) A_{f} = q_{0} {\sum}_{A_f} A_{f}^{1-{\tau}} = 1.0
     \label{normalization}
     \end{equation}
when the sum runs over all clusters.\ \ From eq. (\ref{normalization}) it is obvious that the value of the overall cluster
distribution normalization constant, $q_0$, is dependent on $\tau$ via a Riemann $\zeta$-function:
     \begin{equation}
     q_{0} = 1.0 / {\sum}_{A_f} A_{f}^{1-{\tau}}.
     \label{zeta_function}
     \end{equation}
The above is true only if the scaling assumptions in the FDM apply to all clusters.\ \ For finite size systems even at the
critical point this is only approximately true.\ \ However, it will be seen that eq. (\ref{zeta_function}) holds reasonably well at
the critical point for systems with a continuous phase transition over some range in cluster size.

In the FDM it is assumed that all clusters of size $A_f$ can be treated as an ideal gas, so that the total pressure of the entire
cluster distribution can be determined by summing all of the partial pressures:
     \begin{eqnarray}
     P/(k_{b} T) & = & {\sum}_{A_f} n_{A_{f}}({\epsilon}) \nonumber \\
                 & = & q_{0} {\sum}_{A_f} A_{f}^{-{\tau}} f(z) {g({\mu},T)}^{A_{f}} = M_0 ({\epsilon})
     \label{pressure_a}
     \end{eqnarray}
Is is clear from eq. (\ref{pressure_a}) that the pressure of the system is related to the zeroth moment of the cluster
distribution.

The density is then:
     \begin{eqnarray}
     {\rho} = {\frac{\partial P}{\partial \mu}} & = & q_{0} {\sum}_{A_f} A_{f}^{1-{\tau}} f(z) {g({\mu},T)}^{A_{f}} \nonumber \\
          &   & = {\sum}_{A_f} n_{A_{f}} ({\epsilon}) A_{f} = M_1 ({\epsilon}).
     \label{density}
     \end{eqnarray}
The density is given by the first moment of the cluster distribution.

It is now a simple matter to derive the power law which describes the divergence of the isothermal compressibility,
${\kappa}_{T}$.\ \ By definition:
     \begin{equation}
     {\kappa}_{T} = -
     { \frac{1} {V} } ( { \frac {\partial V} {\partial P} }  )_{T} = 
     { \frac{1} {\rho} } ( { \frac {\partial \rho} {\partial P} } )_{T}.
     \label{compressibility_a}
     \end{equation}
Noting that $k_b T {\rho} = g({\mu},T) ( {\partial} P / {\partial} g({\mu},T))$, eq. (\ref{compressibility_a}) can be rewritten as:
     \begin{eqnarray}
     {\kappa}_T & = & {\frac{-1}{\rho ^2}} \times \nonumber \\
                &   & \left( g(\mu ,T) {\frac{\partial P}{\partial g(\mu ,T)}} + 
                      {g(\mu ,T)}^{2} {\frac{\partial ^2 P}{\partial g(\mu ,T)^2}} \right) _T, 
     \label{compressibility_b}
     \end{eqnarray}
which leads to:
     \begin{eqnarray}
     {\kappa}_{T} & = & ( {\rho} k_{b} T )^{-1} + ( {\rho}^2 k_{b} T )^{-1} {\sum}_{A_f} n_{A_{f}}({\epsilon}) A_{f}^{2} \nonumber \\
                  & = & ( {\rho} k_{b} T )^{-1} + ( {\rho}^2 k_{b} T )^{-1} M_2 ( \epsilon ).
     \label{compressibility_c}
     \end{eqnarray}
The sum in the second term illustrates the relation of the second moment of the cluster distribution, $M_{2}({\epsilon})$, to the 
isothermal compressibility.\ \ The sums in eq. (\ref{pressure_a}), (\ref{density}) and (\ref{compressibility_c}) run over all clusters 
in the gas and exclude the bulk liquid drop.\ \ In percolation and multifragmentation the largest cluster on the liquid side of the 
critical point will be considered as the liquid drop and will thus be excluded from the sum.\ \ On the gas side of the critical point, 
the sum runs over all clusters as there is no longer a liquid drop.

In the thermodynamic limit, large $A_f$ dominate the sum so that it may be treated as an integral giving:
     \begin{equation}
     {\kappa}_{T} = ( {\rho} k_{b} T )^{-1} + ( {\rho}^2 k_{b} T )^{-1} 
     \int_0^{\infty} n_{A_{f}}({\epsilon})  A_{f}^{2} dA_{f}.
     \label{compressibility_d}
     \end{equation}
Working along the liquid-gas coexistence curve so that $g({\mu},T) = 1$ eq. (\ref{compressibility_d}) reduces to:
     \begin{equation}
     {\kappa}_{T} = ( {\rho} k_{b} T)^{-1} + ( {\rho}^2 k_{b} T )^{-1} 
     \int_0^{\infty} A_{f}^{2-{\tau}} f(z) dA_{f}.
     \label{compressibility_e}
     \end{equation}
A change of variables from $A_f$ to $z$ shows that near the critical point:
     \begin{eqnarray}
     {\kappa}_{T} & \sim & ( {\rho}^2 k_{b} T )^{-1} \left | {\frac{q_0}{\sigma}}  \int_0^{\pm \infty} dz 
                           \:f(z) \:|z|^{\frac{3 - \tau - \sigma}{\sigma}} \right | |\epsilon|^{\frac{\tau - 3}{\sigma}} \nonumber \\
                  & = & ( {\rho}^2 k_{b} T )^{-1} {\Gamma}_{\pm} |\epsilon|^{-\gamma} 
     \label{compressibility_f}
     \end{eqnarray}
This is the so-called $\gamma$-power law which describes the divergence of the isothermal compressibility and the second moment of
the cluster distribution near the critical point.\ \ The scaling relation between the exponents $\gamma$, $\sigma$ and $\tau$ is:
     \begin{equation}
     {\gamma} = {\frac{3-{\tau}}{\sigma}} .
     \label{scaling_1}
     \end{equation}

The absolute normalization constants of the $M_{2}({\epsilon})$ power law depend on the scaling function, $f(z)$, the exponent 
$\sigma$ and the overall normalization of the cluster distribution, $q_0$, which in turn depends on the exponent $\tau$:
     \begin{equation}
     {\Gamma}_{\pm} = \left | {\frac{q_0}{\sigma}} 
     \int_0^{\pm \infty} dz \:f(z) \:|z|^{\frac{3 - \tau - \sigma}{\sigma}} \right | .
     \label{compressibility_g}
     \end{equation}
The second moment is related to the isothermal compressibility by the temperature and density of the system.

The derivation of the $\gamma$-power law demonstrates one way to arrive at the scaling relations between the critical exponents.\ \
In addition it illustrates the existence of only two independent exponents and shows the relation of the moments of the cluster
distribution to familiar thermodynamic quantities.\ \ Fisher's framework here illustrated and tempered by percolation theory will
be used in the analysis of the cluster distributions of the three systems discussed above.\ \ It will be seen that in the case of
systems which exhibit a continuous phase transition, the framework of Fisher is well followed, while for systems with no such phase
transition, the framework fails, as it should.

\section{Phase transition signatures in cluster distributions}

\subsection{Insensitive signatures}

In this section the insensitive features of the cluster distribution for each system are examined.\ \ It will be demonstrated that
on this level of analysis each system exhibits behaviors that are consistent with systems which undergo a continuous phase
transition.\ \ The conclusion is inescapable that this sort of analysis can yield necessary, but not sufficient, signals and no 
further insight to the mechanism behind multifragmentation.\ \ A deeper analysis will be necessary to distinguish those systems 
which undergo such a phase transition from those which do not.

\subsubsection{Fluctuations}

One of the most striking characteristics of systems undergoing continuous phase transitions is the occurrence of fluctuations that
exist on all length scales in a small range of the control parameter.\ \ In fluid systems this was observed as critical
opalescence, first noted by Andrews in the latter half of the $19^{th}$ century \cite{theory_of_critical_phenomena}.\ \
Fluctuations in cluster size and the density of the system arise because of the disappearance of the latent heat at the critical
point.\ \ This is illustrated in the FDM when the isothermal compressibility diverges at the critical point and small changes in
pressure gives rise to great changes in the density.\ \ In the FDM as the volume and surface contribution to the free energy of
cluster formation vanishes the power law dominates and clusters of all length scales are observed \cite{fisher}.

In a cluster distribution the most readily observed fluctuations are those in the size of the largest cluster.\ \ For each system
the root mean square (RMS) fluctuations in the size of the largest cluster normalized to the size of the system, ${\Delta} \left(
A_{max} / A_{0} \right)$, have been calculated as a function of the system's control parameter.\ \ This measure of the
fluctuations in the cluster distribution was first studied by Campi for gold multifragmentation and percolation \cite{campi_2}.\ \
Those results are replicated here for those two systems.

Figure 1a shows ${\Delta} \left( A_{max} / A_{0} \right)$, as a function of $p_l$ for percolation.\ \ As expected for a system
known to exhibit a continuous phase transition, the RMS fluctuations peak over a narrow range in the control parameter.\ \ The
location of this peak provides a first estimate of the critical point; $p_c = 0.33 \pm 0.01$.\ \ See Table I.

\begin{figure} [h]
\centerline{\psfig{file=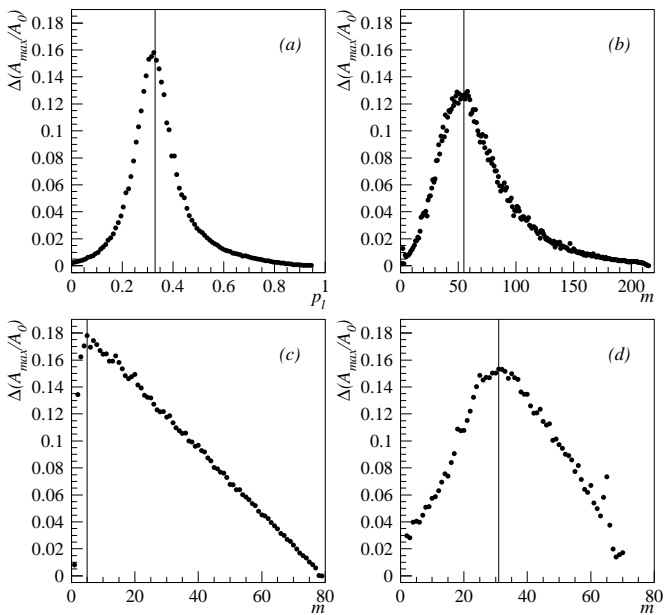,width=10.0cm,angle=0}}
\caption{Fluctuations in the normalized size of the largest cluster for: (a) percolation ($A_0 = 216$) as a function of $p_l$,
(b) percolation as a function of $m$, (c) random partitions as a function of $m$ and (d) Au $+$ C multifragmentation as a function
of $m$.\ \ Solid vertical lines indicate the guess for the critical point.}
\label{fig:01}
\end{figure}

\begin{table*}
\begin{center}
\caption{Critical point determination}
\begin{tabular}{lllll}
\hline
\multicolumn{1}{} 
.Method $/$ System                         & Percolation ($p_l$) & Percolation ($m$) & Random Partitions & Au $+$ C          \\
\hline
${\Delta} \left( A_{max} / A_{0} \right)$ &  $0.33 {\pm} 0.01$  &  $55 {\pm} 5 $    &   $ 5 {\pm} 2$    &   $31 {\pm} 6$    \\
$m_2$ peak                                &  $0.28 {\pm} 0.03$  &  $62 {\pm} 2 $    &   $ 5 {\pm} 2$    &   $18 {\pm} 2$    \\
Fisher $\tau$-power law                   &  $0.31 {\pm} 0.05$  &  $57 {\pm} 3 $    &   $59 {\pm} 1$    &   $22 {\pm} 1$    \\
Scaling Function                          &  $0.34 {\pm} 0.03$  &  $57 {\pm} 6 $    &   $10 {\pm} 2$    &   $22 {\pm} 2$    \\
$\gamma$-matching                         &  $0.33 {\pm} 0.02$  &  $49 {\pm} 1 $    &   $ 9 {\pm} 1$    &   $21 {\pm} 2$    \\
\hline
\end{tabular}
\end{center}
\end{table*}

Next the percolation lattice is examined using the multiplicity of clusters, $m$, as an estimate of the control parameter.\ \ This
is done because in the case of nuclear multifragmentation $m$ is experimentally measurable.\ \ Figure 1b shows much the same 
qualitative behavior as Figure 1a.\ \  The fluctuations peak over some narrow range of $m$ and suggest the value of the multiplicity 
at the critical point, the {\it critical multiplicity}, to be $m_c = 55 \pm 5$.

For random partitions a peaking behavior in the fluctuations of the size of the largest cluster as a function of $m$ was observed,
see Figure 1c.\ \ These fluctuations can be understood as follows.\ \ At $m = 1$ there can be no fluctuations in the size of the
largest cluster because of the dual constraints of event cluster multiplicity and the fixed number of constituents.\ \ As the
multiplicity increases from one, the constraints ease and fluctuations in the size of the largest cluster grow.\ \ At the maximum 
possible multiplicity, {\it i.e.} when $m$ is equal to the total number of constituents, the size of the largest cluster is 
constrained to be equal to unity.\ \ Thus, the fluctuations show a peak, but for a reason that has nothing to do with a continuous 
phase transition.\ \ Therefore it must be concluded that the observation of a maximum in the fluctuations in the size of the largest 
cluster is not sufficient to distinguish systems with and without critical behavior.\ \ On the other hand, the absence of a peak in 
fluctuations would indicate that the clusters of the system were not produced near a critical point.\ \ If the system's phase space 
has been fully explored, then the stronger statement that the system does not possess a critical point could be made.\ \ At this 
level of analysis the critical multiplicity of this system can be estimated to be $m_c = 5 \pm 2$.

Finally, Figure 1d shows the Au $+$ C multifragmentation data with the cluster distribution normalized to the size of the system,
$A_0(m)$.\ \ The fluctuations in the mass of the largest cluster exhibit a peak when plotted as a function of the event total
charged particle multiplicity, $m$.\ \ This behavior is consistent with what is expected for a critical phenomenon.\ \ However, as
illustrated above, it is far from conclusive.\ \ At this level of analysis the estimate for the critical multiplicity is $m_c =
31 \pm 6$.

It is also possible to study the fluctuations in the average size of a cluster.\ \ From the example of critical opalescence it is
clear that the greatest fluctuations in cluster size should occur at the critical point.\ \ To  that end the quantity known as
${\gamma}_2$ is constructed again following the work of Campi \cite{campi_2} - \cite{campi_7}.\ \ The variance
in the mean cluster size, $\left<{A_f}\right>$, is defined as:
     \begin{equation} 
     {\sigma}^2 = { \lim_{N \rightarrow \infty} ( {\frac{1}{N}}
     \sum A_{f}^{2} ) - {\left<{A_f}\right>}^2 } . 
     \label{gamma_2_a} 
     \end{equation}
The average cluster size is given by the ratio of the first moment to the zeroth moment:
     \begin{equation} 
     \left<{A_f}\right> = \sum n_{A_f} A_f / \sum n_{A_f} = M_1 / M_0.
     \label{gamma_2_b} 
     \end{equation}
The first term in eq. (\ref{gamma_2_a}) is just the ratio of the second moment to the zeroth moment.\ \ Therefore, the variance in
the average cluster size can be written in terms of the $k^{th}$-moments:
     \begin{equation} 
     {\sigma}^2 = {\frac{M_2}{M_0}} - ({\frac{M_1}{M_0}})^2 .
     \label{gamma_2_c} 
     \end{equation}
This quantity is directly related to Campi's ${\gamma}_2$ via:
     \begin{equation} 
     {\gamma}_2 = {\frac{{\sigma}^2}{\left<A\right>^2}} + 1 = {\frac{M_2 M_0}{M_{1}^2}} , 
     \label{gamma_2_d} 
     \end{equation}
which is easily measured and was coined by Campi as the {\it reduced variance} \cite{campi_2}.

In a later paper, \cite{campi_6}, Campi discussed the differences in methods to measure ${\gamma}_2$.\ \ Specifically, the
manner in which the $k^{th}$-moments are computed from the observed cluster distribution.\ \ One method is to measure the
$k^{th}$-moments on an event by event basis and then compute an average based on the control parameter, {\it e.g.}:
     \begin{equation}
     \left<M_k ({\epsilon})\right> = {\frac{1}{N}} \sum_{i=1}^{N} M_{k}^{i} ({\epsilon}) =
     {\frac{1}{N}} \sum_{i=1}^{N} ( \sum_{A_f} n_{A_{f}}^{i} ({\epsilon}) A_{f}^{k} )
     ,
     \label{ave_sum} 
     \end{equation}
where $N$ is the number of events at a control parameter value of $\epsilon$, and $i$ denotes the $i^{th}$ event.\ \ This method of
calculation of the $k^{th}$-moments will be termed {\it averaging the sums} and will yield: $\left<{{\gamma}_2}\right>$. 

The alternate method involves calculating an average cluster distribution at each value of the control parameter and then
calculating the $k^{th}$-moments from the resulting average cluster distribution:
     \begin{equation} 
     {\overline{M}}_{k} = \sum \left<{n_{A_f}(\epsilon)}\right> A_{f}^k = \sum_{A_f}
     ({\frac{1}{N}} {\sum_{i=1}^{N}} n_{A_f}^i (\epsilon)) A_{f}^k .
     \label{sum_ave_a} 
     \end{equation}
This method of calculation will be termed {\it summing the averages} and will give: ${\overline{\gamma}}_{2}$.

For quantities linear in $n_{A_f}$ there is no difference in the two methods so that $\left<{M_{k}( \epsilon )}\right> =
{\overline{M}}_{k}( \epsilon )$.\ \ However, due to the dependence of ${\gamma}_2$ on the square of the first moment, there will be
a difference in the two methods of calculation.\ \ Results for both methods for each system are shown in Figure 2.

\begin{figure} [h]
\centerline{\psfig{file=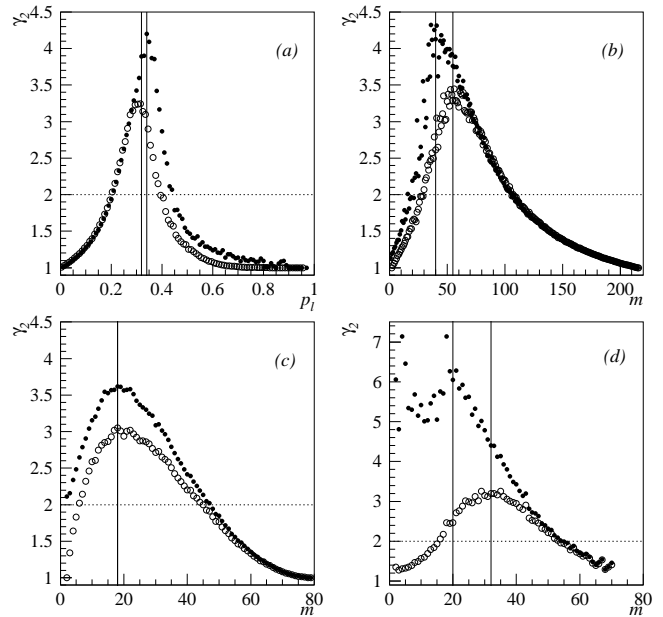,width=10.0cm,angle=0}}
\caption{Fluctuations as measured by ${\gamma}_2$ for: (a) percolation as a function of $p_l$, (b) percolation as a function of
$m$, (c) random partitions as a function of $m$ and (d) Au $+$ C multifragmentation as a function of $m$ (at low $m$ 
${\overline{{\gamma}}_{2}}$ is affected by fission events not completely filtered).\ \ Open circles show $\left<{\gamma}_{2}\right>$, 
while filled circles show ${\overline{{\gamma}}_{2}}$ (see text for details).\ \ Solid vertical lines indicate the guess for the 
critical point.\ \ A dotted horizontal line shows the value of ${\gamma}_{2} = 2$.}
\label{fig:02}
\end{figure}

Of primary significance is the presence of a peak in both measurements of ${\gamma}_2$ for all systems.\ \ For an infinite system
exhibiting critical phenomena, the location of the peak in ${\gamma}_2$ will coincide with the location of the critical point.\ \
For the percolation system Figures 2a and 2b show that both the location and magnitude of the peak in ${\gamma}_2$ is dependent on
the choice of calculation method.\ \ Solid lines indicate this measure of the critical point.\ \ For the random partitions, Figure
2c, the location of the peak in ${\gamma}_2$ shows no dependence on the method of calculation while the magnitude of the peak
does.\ \  The gold multifragmentation data exhibit a dependence on the method of calculation both in the magnitude and location of
a peak in ${\gamma}_2$.

Having noted the peaking behavior of ${\gamma}_2$, the significance of the amplitude of the peak is now addressed.\ \ It has been
suggested that the height of the peak can be used to differentiate between the presence of a power law and that of an exponential:
for a power law ${\gamma}_2 > 2$ while for an exponential ${\gamma}_2 < 2$.\ \ This is not definitive proof of the existence of a
continuous phase transition as other systems show power laws in the absence of such a phase transition.\ \ All of the percolation
figures show peaks above two, as do the multifragmentation data plots and the random partitions.\ \ However, the value of
${\gamma}_2$ depends on the size of the system in question \cite{campi_6}.\ \ For a percolation system with 64 sites, peaks
in ${\gamma}_2$ under two are observed, see Figure 3a and 3c.\ \ Therefore, the criterion ${\gamma}_2 > 2$ is not sufficient to
discriminate between those finite systems which do and those which do not posses a power law cluster distribution.

\begin{figure} [h]
\centerline{\psfig{file=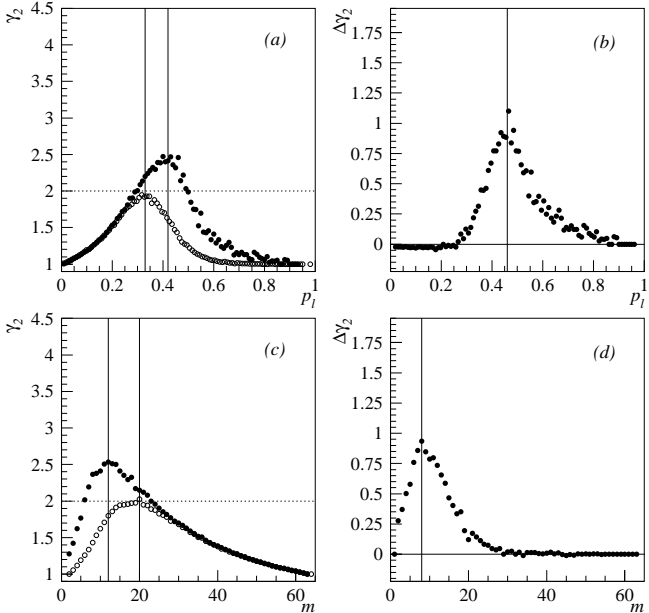,width=10.0cm,angle=0}}
\caption{Fluctuations as measured by ${\gamma}_2$ (see text for details) for a percolation lattice of 64 sites: (a)open circles
$\left<{\gamma}_{2}\right>$, filled circles ${\overline{{\gamma}}_{2}}$ as a function of $p_l$ , (b)${\Delta}{\gamma}_{2}$ as a
function of $p_l$, (c) open circles $\left<{\gamma}_{2}\right>$, filled circles ${\overline{{\gamma}}_{2}}$ as a function of $m$
and (d) ${\Delta}{\gamma}_{2}$ as a function of $m$.\ \ Solid vertical lines indicate the guess for the critical point.\ \ A dotted
horizontal line shows the value of ${\gamma}_{2} = 2$.}
\label{fig:03}
\end{figure}

Finally the question of the difference between the alternative methods of calculating ${\gamma}_2$  is examined via: $\Delta
{\gamma}_2 = \left< {\gamma}_2 \right> - {\overline{\gamma}}_{2}$.\ \ It has been suggested that a peak in the difference could
indicate critical phenomena and the location of the critical point \cite{campi_6}.\ \ Unfortunately, the cause of this peak is not
well understood and vanishes at the limits of the system size: $(0, \infty )$.\ \ Figures 4a and 4b do show peaks in $\Delta
{\gamma}_2$ at some intermediate value of the control parameter for this percolation lattice of 216 sites.\ \ However, as the size
of the percolation lattice increases this signal vanishes \cite{campi_6}.\ \ For a percolation lattice with 64 sites Figures 3c and
3d, respectively,  look like a cross between the percolation ($L = 6$, $m$) results, Figures 2b and 4b, and the random partition
results shown in Figures 2c and 4c.\ \ This is believed to be due to the twin constraints of the multiplicity and the conservation
of constituents imposed upon the system at the extremes in cluster multiplicity.\ \ Similar behavior is observed in the gold
multifragmentation data in Figures 2d and 4d.

\begin{figure} [h]
\centerline{\psfig{file=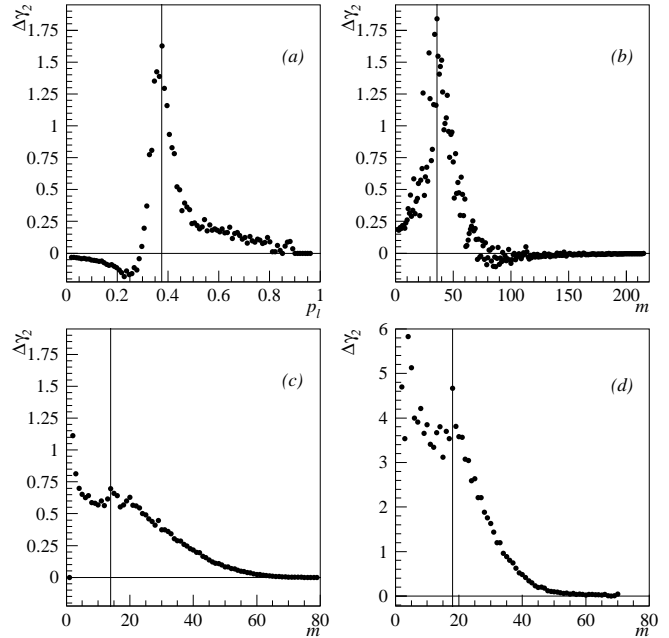,width=10.0cm,angle=0}}
\caption{The quantity ${\Delta}{\gamma}_{2}$ for: (a) percolation ($A_0 = 216$) as a function of $p_l$, (b) percolation as a
function of $m$, (c) random partitions as a function of $m$ and (d) Au $+$ C multifragmentation as a function of $m$.\ \ Solid
vertical lines indicate the guess for the critical point.}
\label{fig:04}
\end{figure}

Neither the ${\gamma}_2$ measure of fluctuations nor the observation of fluctuations in the size of the largest cluster provide
definitive insight into the nature of the cluster producing mechanism.\ \  For both random partitions and percolation ${\gamma}_2$ 
peaks at nearly the same value of the control parameter regardless of the method of averaging used.\ \ For the percolation system 
the value of $p_l$ at the peak in ${\gamma}_2$ is close to  the value of $p_l$ where ${\Delta}(A_{max}/A_0)$ is a maximum.\ \ This 
coincidence does not hold for random partitions; compare Figure 1c and 4c.\ \ For both percolation ($m$) and multifragmentation, 
there is better agreement on the critical point from fluctuations and than from ${\gamma}_2$ is computed via eq. (\ref{ave_sum}).

\subsubsection{Divergences}

Another signature previously used to infer the existence of a continuous phase transition from cluster distributions is the
observance of a peak in the second moment \cite{campi_1}, \cite{bonesara}.\ \ It has been pointed out that models with no phase
transition can exhibit a peaking behavior in the second moment \cite{phair}.\ \ Figure 5 shows the behavior of the second moment
for each of the systems examined in this work.\ \ In this figure, for the sake of illustration, the largest cluster has been 
excluded from the sum at all values of the control parameter.\ \ Each system shows a peak at some intermediate value of its control 
parameter.\ \ Table I lists the location of the second moment peaks.\ \ It is clear from the peak observed for the random partitions 
that it is possible to observe a peak in the second moment for a non-critical cluster distribution.\ \ Thus this quantity cannot be 
used to distinguish between critical and non-critical systems.

\begin{figure} [h]
\centerline{\psfig{file=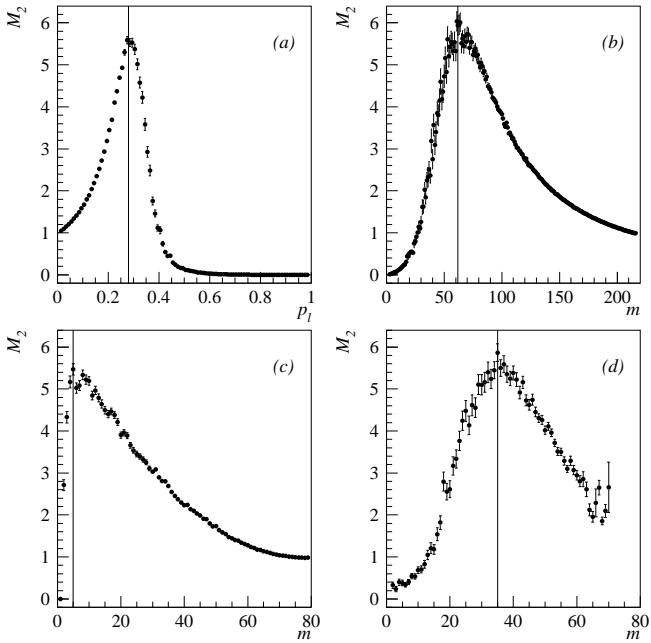,width=10.0cm,angle=0}}
\caption{The second moment, $M_{2}$, of the cluster distributions for: (a) percolation as a function of $p_l$, (b) percolation
as a function of $m$, (c) random partitions as a function of $m$ and (d) Au $+$ C multifragmentation as a function of $m$.\ \ Solid
vertical lines indicate the guess for the critical point.}
\label{fig:05}
\end{figure}

An issue with the use of the second moment's peaking behavior is the exclusion of the largest cluster from the sum in eq.
(\ref{compressibility_c}).\ \ Again, in the FDM formalism the sum runs over all clusters in the {\it gas}.\ \ On the {\it liquid} 
side of the critical point a gas exists in addition to a liquid drop.\ \ Thus, the largest cluster represents the bulk liquid.\ \ 
On the {\it gas} side of the critical point there is no liquid drop and the largest cluster is merely the largest gas particle.\ \ 
With this understanding it is clear that the largest cluster should be omitted from the summation in the second moment only in the 
{\it liquid} region, whereas the summation should run over all clusters in the {\it gas} region.\ \ For a proper construction of the
second moment, knowledge of the location of the critical point is required.\ \ In the thermodynamic limit of infinite system size,
exclusion of the largest cluster makes little difference.\ \ However in small systems the proper construction of the second moment
is crucial if critical behavior is to be observed in ref. \cite{elliott_perc_1}.\ \ See ref. \cite{mastinu} for an example of the 
improper construction of the second moment.

\subsubsection{Campi plots}

Plots of the natural log of the normalized size of the largest cluster, $\ln ( A_{max} / A_0 )$, versus the natural log of the
second moment, $\ln ( M_2 )$, were first presented by Campi in a comparison of gold multifragmentation and percolation
\cite{campi_1}.\ \ Figure 6 shows the resulting plots for each of the systems discussed in this paper.\ \ In each plot there is a
{\it liquid leg} for the largest $A_{max}$ and small $M_2$ and a {\it gas leg} for smaller $A_{max}$ and mid-range values of
$M_2$.\ \ That similar behavior is observed for all systems is in indication that this is a necessary, but not sufficient, signal 
for critical behavior.

\begin{figure} [h]
\centerline{\psfig{file=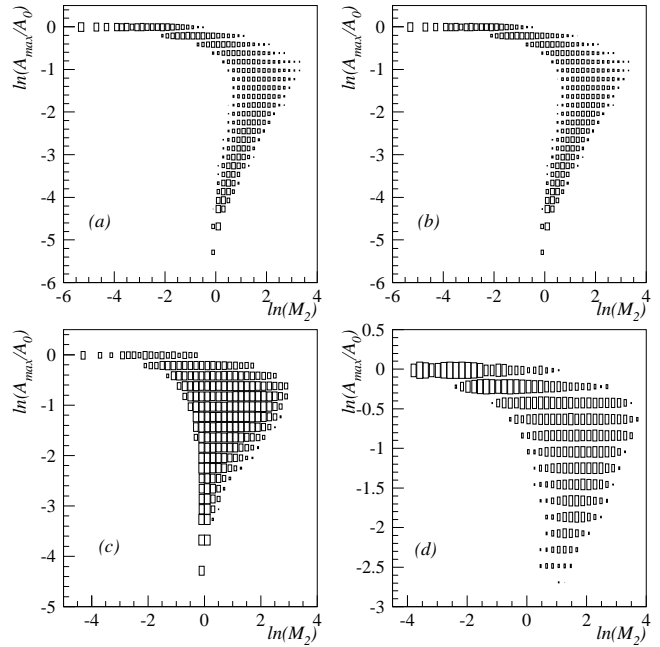,width=10.0cm,angle=0}}
\caption{Campi plots of the cluster distributions for: (a) percolation as a function of $p_l$, (b) percolation as a function
of $m$, (c) random partitions as a function of $m$ and (d) Au $+$ C multifragmentation as a function of $m$.}
\label{fig:06}
\end{figure}

\subsubsection{Rise and fall of intermediate mass fragments}

In many nuclear multifragmentation studies the term {\it intermediate mass fragment}, IMF, has been defined as a cluster which has a
charge between $3 \le Z_f \le 30$.\ \ For the percolation system presented here a {\it charge} has been assigned to each cluster by
multiplying the number of constituents in the cluster by the charge to mass ratio of a gold nucleus.\ \ For the random partitions
the number of constituents is used as the {\it charge}.\ \ Since the definition of an IMF is arbitrary it makes little qualitative 
difference what range in some measure of the cluster size is used.

Aside from the equilibrium arguments made by the ALIDIN group \cite{gsi_1}-\cite{gsi_3}, little insight towards the presence or 
absence of a continuous phase transition is gained from a simple plot of the average number of IMF's, $\left<M_{imf}\right>$ versus 
the control parameter.\ \ Figure 7 shows the results for the systems discussed in this work.\ \ Each system shows a peak in 
$\left<M_{imf}\right>$ at some intermediate value of the control parameter.\ \ Comparing the peak position in Figure 7 to the values 
listed in Table I shows that there is little correspondence between the numerous proposed methods for locating the critical point.\ \ 
The arbitrary nature of the definition of an IMF makes it unlikely that the peak in $\left<M_{imf}\right>$ occurs at the critical
point.\ \ To some degree the rise and  fall feature is due to the constraint of a fixed number of constituents.\ \ It it obvious
that a the extreme values of the control parameter, the number of IMFs must diminish, while at intermediate values, it must be at
least as great. \ \  Thus, the occurrence of a peak at some intermediate value of the control parameter is expected.

\begin{figure} [h]
\centerline{\psfig{file=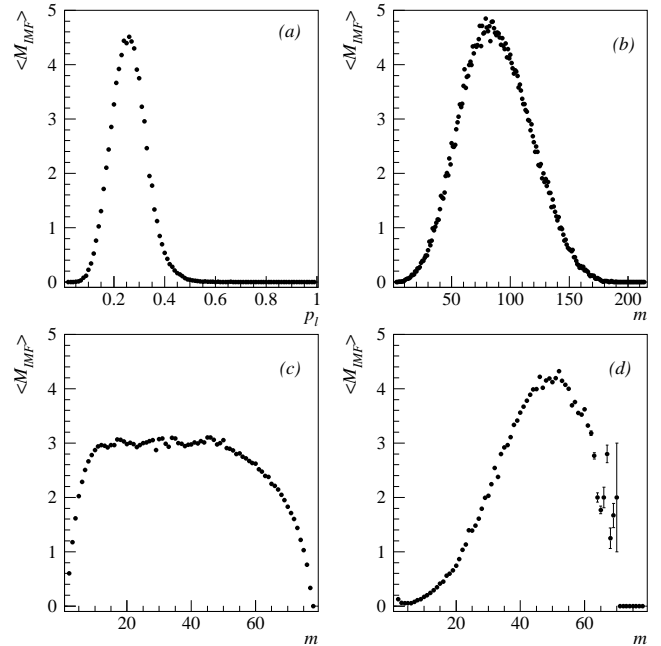,width=10.0cm,angle=0}}
\caption{Average number of intermediate mass clusters, $\left<M_{IMF}\right>$, for: (a) percolation as a function of $p_l$, (b)
percolation as a function of $m$, (c) random partitions as a function of $m$ and (d) Au $+$ C multifragmentation as a function of
$m$.}
\label{fig:07}
\end{figure}

\subsubsection{${\tau}_{eff}$-minimum}

With the first observation of a power law in the nuclear multifragmentation yield distribution \cite{finn}, \cite{minich} it became
a common analysis tool to fit cluster distributions to a power laws and extract exponent values.\ \ In an effort to make this a
more quantitative analysis the value of the extracted exponent, ${\tau}_{eff}$, was examined as a function of some  control
parameter that was experimentally or numerically accessible.\ \  It was assumed that at the critical point the value of
${\tau}_{eff}$ should attain a lower value than fits which were performed away from the critical point
\cite{panagiotou}-\cite{williams}.\ \ The logic of this assumption was based upon the idea that at low {\it temperatures} a system
has few small clusters, so the power law should be steep, leading to a high ${\tau}_{eff}$ value.\ \ At high temperatures there are
many small clusters and little else, which is reflected in a high value of ${\tau}_{eff}$ and a steep power law.\ \ At the critical
point clusters on all length scales appear and the power law is shallow with a lower value of ${\tau}_{eff}$.\ \ In this analysis
the largest cluster was generally omitted from the fitting procedure and both the constant of proportionality and ${\tau}_{eff}$
were allowed to vary independently.\ \ Many investigations of nuclear multifragmentation, both theoretical and experimental,
employed this method of analysis \cite{panagiotou} -\cite{williams}.

There are two flaws in this analysis method.\ \ The first is the use of a two parameter fit for the power law.\ \ Allowing
both the overall normalization of the power law and the exponent to vary independently is in conflict to the scaling
assumptions underlying the FDM as shown in eq.'s (\ref{normalization}) and (\ref{zeta_function}).\ \ A proper fit for a power law
within the context of the FDM should be based on single parameter.\ \ Furthermore, the cluster distribution must be normalized to the
size of the system as was outlined in III.\ \ Without this normalization, which requires knowledge of the system's size, power law
fits lose much of their ability to contribute useful information to the presence of critical phenomena.

Leaving aside for a moment that the execution of the ${\tau}_{eff}$-minimum analysis violates the scaling assumptions of the FDM,
the signal of a minimum in the cluster yield power law will be examined.\ \  A two parameter fit for ${\tau}_{eff}$ searches for
the minimum in an effective exponent which is defined as \cite{margolina} -\cite{wfjm_catania}:
     \begin{equation}
     {\tau}_{eff} = - {\frac{ \partial \ln n_{A_f} ( \epsilon ) }{ \partial \ln
     A_f }} .
     \label{tau_eff_a}
     \end{equation}
If it is assumed that the system under study follows a power law in the cluster yield at the critical point, and away from the
critical point the cluster yield is affected by a scaling function such as in eq. (\ref{power_law}), then:
     \begin{equation}
     {\tau}_{eff} = \tau - A_f {\frac{ \partial \ln f }{ \partial A_f }} .
     \label{tau_eff_b}
     \end{equation}
The minimum in ${\tau}_{eff}$ can be found by differentiating eq. (\ref{tau_eff_b}):
     \begin{equation}
     {\frac{ d {\tau}_{eff} }{ d {\epsilon} }} = - A_f {\frac{ \partial }{
     \partial \epsilon }} {\frac{ \partial \ln f }{ \partial A_f }} = 0 .
     \label{tau_eff_c}
     \end{equation} 
This indicates that the location of the minimum in ${\tau}_{eff}$ is dependent on the form of the scaling function, $f$.\ \
Assuming the scaling function has the form of eq.(\ref{surface_term_leath}) then the minimum in ${\tau}_{eff}$ will be at $
\epsilon = B / 2 A_{f}^{\sigma} $ , and not at the critical point $ {\epsilon}_c = 0 $.

Despite the flaws in the ${\tau}_{eff}$-minimum analysis it is of interest to examine the results for the systems discussed
in this paper.\ \ Figures 8 through 11 show the results for a two parameter fit to the cluster distribution for percolation 
(probability and multiplicity), random partitions and gold multifragmentation, respectively.\ \ For all systems, the cluster 
distributions were fit at each value of the control parameter.\ \ Only clusters with ${\sim}0.02 \le ( A_f / A_0 ) \le {\sim}0.22$ 
were included in the fits.\ \ The first three systems weighted ${\chi}_{\nu}^{2}$ with errors associated with $n_{A_f}({\epsilon})$ 
while the ${\chi}_{\nu}^{2}$ for the gold multifragmentation cluster distributions were weighted with errors on both 
$n_{A_f}({\epsilon})$ and $A_f$.

For the percolation ($p_l$) a minimum in ${\tau}_{eff}$ was observed at $p_l = 0.3$ with ${\chi}_{\nu}^{2} = 2.3$, $q_0 = 0.214 \pm
0.005$ and $\tau = 2.19 \pm 0.01$; shown in Figure 8a, b and c by the dotted lines.\ \ However, at $p_l = 0.33$ the
${\chi}_{\nu}^{2} = 1.02$ , $q_0 = 0.181 \pm 0.003$ and $\tau = 2.27 \pm 0.01$; shown in Figure 8a, b and c with the dashed
lines.\ \  Based on a goodness of fit comparison, the latter value of $p_l$ is a better choice for the critical point as the
cluster distribution is better fit by a power law.\ \ This result is in agreement with the analytic discussion of ${\tau}_{eff}$
above, namely that a minimum in ${\tau}_{eff}$ is a poor indicator of the critical point.\ \ If the results for $p_l = 0.33$ are
compared to the center of the ${\tau}_{eff}$, $p_l \sim 0.28$, the differences in the ${\chi}_{\nu}^{2}$ and $q_0$ results are even
more striking.

\begin{figure} [h]
\centerline{\psfig{file=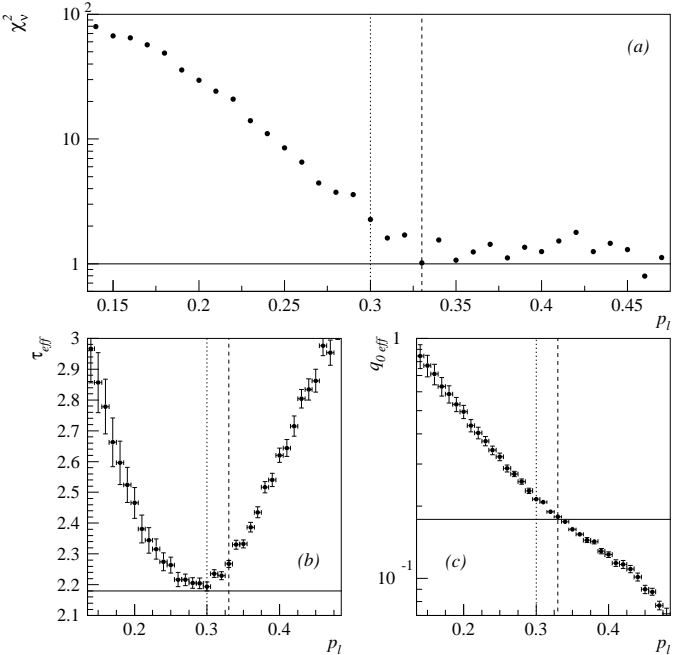,width=10.0cm,angle=0}}
\caption{Results for ${\tau}_{eff}$-minimum analysis on percolation as a function of $p_l$: (a) ${\chi}_{\nu}^{2}$, (b)
fitted ${\tau}_{eff}$ and (c) $q_0$.\ \ The vertical dotted line show the location of the minimum in ${\tau}_{eff}$.\ \ The
vertical dashed line shows one instance of a better fit based on ${\chi}_{\nu}^{2}$.\ \ The horizontal solid lines show the
accepted values of $\tau$ and $q_0$ for percolation.}
\label{fig:08}
\end{figure}

Similar results were seen for percolation ($m$), see Figure 9.\ \ Here the minimum in the ${\tau}_{eff}$-well yielded worse results
for both ${\chi}_{\nu}^{2}$ and $q_0$ than does the choice of the critical point based on a choice from the ${\chi}_{\nu}^{2} \sim
1$ region where there is good agreement between the fitted $q_0$ and the value computed using eq. (\ref{zeta_function}) and the
canonical $\tau$ value for three dimensional percolation.

\begin{figure} [h]
\centerline{\psfig{file=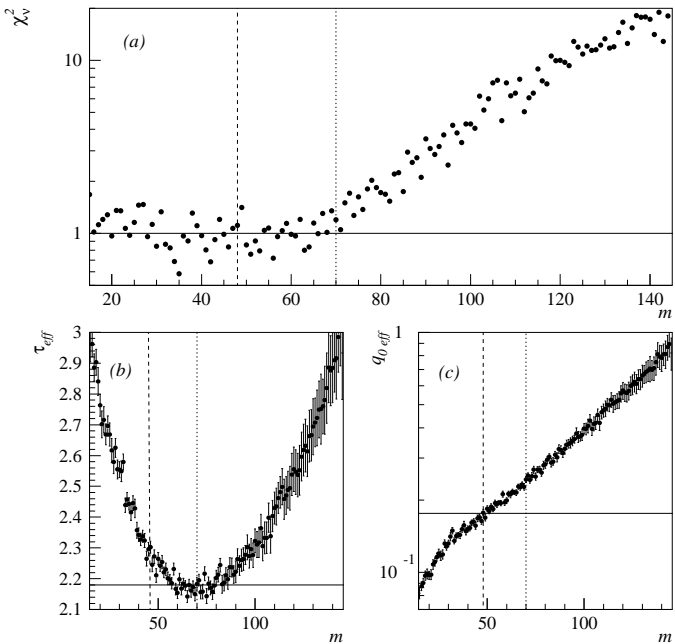,width=10.0cm,angle=0}}
\caption{Results for ${\tau}_{eff}$-minimum analysis on percolation as a function of $m$: (a) ${\chi}_{\nu}^{2}$, (b)
fitted ${\tau}_{eff}$ and (c) $q_0$.\ \ The vertical dotted line show the location of the minimum in ${\tau}_{eff}$.\ \ The
vertical dashed line shows one instance of of a better fit based on ${\chi}_{\nu}^{2}$.\ \ The horizontal solid lines show the
accepted values of $\tau$ and $q_0$ for percolation.}
\label{fig:09}
\end{figure}

Significant differences between percolation and random partitions are observed in this analysis as seen in Figure 10.\ \ The solid
lines show the ${\tau}_{eff}$ and $q_0$ values for systems in the three dimensional Ising universality class, while the dashed line
shows the ${\tau}_{eff}$ and $q_0$ for three dimensional percolation.\ \ The first noticeable difference is a lack of a valley
shape in the plot of ${\tau}_{eff}$ versus control parameter, see Figure 10b.\ \ The value of ${\tau}_{eff}$ is below $2.2$ for all
but $m > 60$.\ \ Next is the lack of a region in $m$ where ${\chi}_{\nu}^{2} \sim 1$ (other than at $m = 2$), Figure 10a.\ \ All
fits yield large ${\chi}_{\nu}^{2}$ values indicating poor fits to the cluster distribution by a power law for the range of
clusters examined.

\begin{figure} [h]
\centerline{\psfig{file=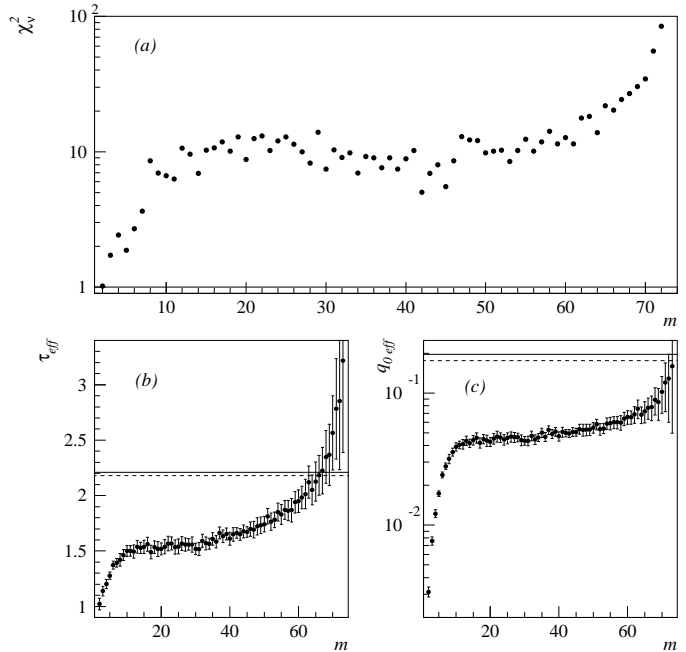,width=10.0cm,angle=0}}
\caption{Results for ${\tau}_{eff}$-minimum analysis on random partitions: (a) ${\chi}_{\nu}^{2}$, (b) fitted ${\tau}_{eff}$
and (c) $q_0$.\ \ The horizontal solid (dashed) lines show the accepted values of $\tau$ and $q_0$ for percolation (3D Ising).}
\label{fig:10}
\end{figure}

The gold multifragmentation data show results similar to those of percolation ($m$).\ \ Here the cluster size is measured in terms
of the nucleon number and the cluster distribution is normalized to the mass of the gold projectile remnant.\ \ Figure 11b shows a
valley in ${\tau}_{eff}$ as a function of $m$, albeit one with a shallow and questionable upwards slope at high $m$.\ \ Figure 11c
shows fitted values of $q_0$ that coincide with canonical values.\ \ Figure 11a shows a region of low ${\chi}_{\nu}^{2}$ values
followed by steadily increasing values.\ \ If no knowledge of the $q_0$ and $\tau$ values is assumed, then this analysis
shows no definitive signals.\ \ The ${\tau}_{eff}$-valley shows a broad minimum in ${\chi}_{\nu}^{2}$ thus no one value of $m$ can
be selected for the critical point based on goodness of fit arguments.\ \ At best one could argue for the neighborhood of the
critical point and a value of $q_0$ and $\tau$ in some broad range.

\begin{figure} [h]
\centerline{\psfig{file=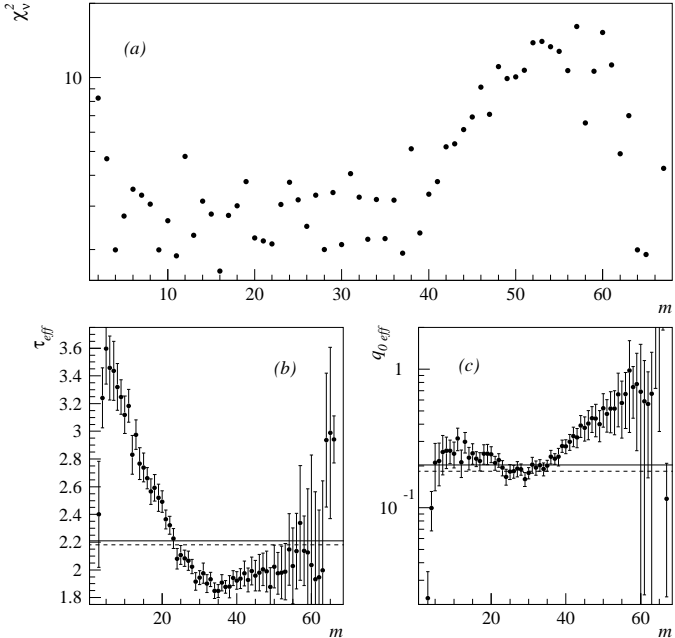,width=10.0cm,angle=0}}
\caption{Results for ${\tau}_{eff}$-minimum analysis on Au $+$ C multifragmentation: (a) ${\chi}_{\nu}^{2}$, (b) fitted
${\tau}_{eff}$ and (c) $q_0$.\ \ The horizontal solid (dashed) lines show the accepted values of $\tau$ and $q_0$ for percolation
(3D Ising).}
\label{fig:11}
\end{figure}

\subsubsection{Conclusion}

The analysis presented in this section shows that many proposed indicators of critical behavior are inconclusive.\ \ All of the
considered systems show similar signals which are qualitative in nature and open to interpretation.\ \ It is therefore impossible, 
based solely on this level of analysis, to make a definitive conclusion as to the presence of a continuous phase transition in any 
of these systems.\ \ What is needed is an analysis or set of analyses that more clearly differentiates between systems with and without 
critical behavior.

\subsection{Sensitive signatures}

\subsubsection{The Fisher $\tau$-power law and the critical point}

In this section the cluster yields are fit to a power law in a manner consistent with the FDM formalism.\ \ As with the two
parameter fits the yields for clusters with ${\sim}0.02 \le ( A_f / A_0 ) \le {\sim}0.22 $ were fit at each value of the control
parameter.\ \ However, only a single parameter, the value of $\tau$, was allowed to vary to minimize the ${\chi}_{\nu}^{2}$ of the
fit.\ \ The value of the normalization, $q_0$, was determined via the Riemann $\zeta$-function in eq. (\ref{zeta_function}).\ \ As
suggested by Fisher \cite{fisher}, the value of $\tau$ was constrained to be between 2 and 3 so that the sum in the $\zeta$-function
converges.

If the cluster distribution is well described by the FDM, then at the critical point the fit to a single parameter power law
should show a minimum in ${\chi}_{\nu}^{2}$.\ \ Away from the critical point the power law is modified by a scaling function with a
form similar to that given in eq. (\ref{surface_term_leath}).\ \ Therefore, fits to a single parameter power law should become
worse as the modification from the scaling function increases away from the critical point.

Figure 12 shows the results for the percolation system with $p_l$ as the control parameter.\ \ In Figure 12a, a minimum in
${\chi}_{\nu}^{2}$ is observed for fits in the mid $p_l$ range.\ \ This minimum indicates the location in $p_l$ of the cluster yield 
distribution which is best fit by a single parameter power law as suggested by the FDM formalism.\ \ By this estimation the critical 
point for this 216 site percolation lattice is at $p_c = 0.31 \pm 0.05$ with $\tau = 2.2 \pm 0.1$, $q_0 = 0.20 \pm 0.01$ and a 
${\chi}_{\nu}^{2} = 1.62$.\ \ The canonical values of $\tau$ and $q_0$ are not extracted due in part to unavoidable finite size 
effects, and to the binning of cluster yields together over a range of 0.01 in $p_l$, which causes the {\it true} cluster distribution 
at the critical point to be contaminated by distributions at other values of the control parameter.\ \ In spite of these difficulties, 
the signature of the critical behavior suggested by the FDM formalism is unmistakable.\ \ The location of the critical point determined 
here is consistent, at the 10$\%$ level, with the insensitive signatures presented in the previous section.\ \ See Table I.\ \ Figure 
16a shows the best fit power law.\ \ For the percolation system clusters consisting of a single site are excluded from the fitting 
procedure.\ \ It is accepted that those clusters reflect the effects of the finite size of the system to a higher degree than larger 
clusters.\ \ Clusters with $ A_f \le 53$ were included in the fit.\ \ The largest cluster from each {\it event} was excluded from 
consideration when generating the average cluster distribution in keeping with the FDM formalism.\ \ Figure 16a shows the data for the 
entire cluster distribution in open circles.\ \ It is clear from this figure that the majority of the cluster distribution was used in 
the power law fit and further, that the exclusion or inclusion of the larger clusters has almost no effect on the results of this 
procedure.\ \ In short, the extracted parameters, namely $\tau$, $q_0$ and $p_c$, do not depend on the fit range.

\begin{figure} [h]
\centerline{\psfig{file=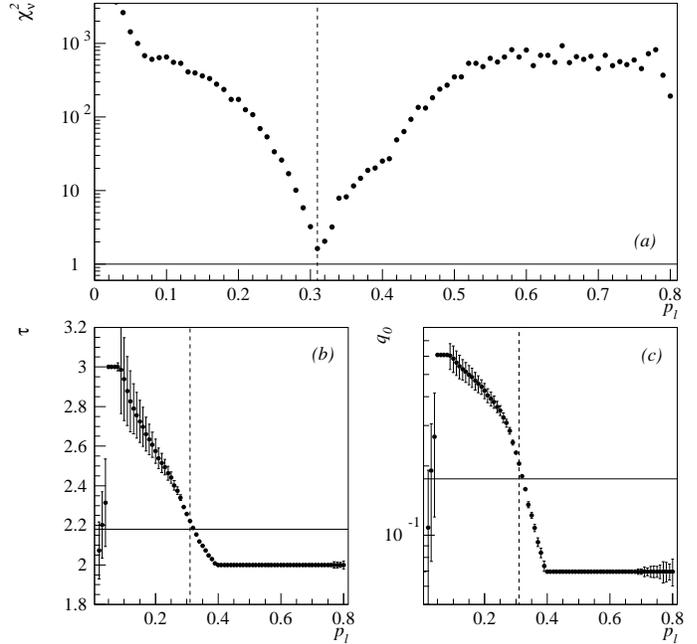,width=10.0cm,angle=0}}
\caption{Results for the full Fisher $\tau$-power law analysis on percolation as a function of $p_l$: (a) ${\chi}_{\nu}^{2}$,
(b) fitted ${\tau}$ and (c) $q_0$.\ \ The vertical dashed line shows the estimate of the critical point based on a best fit
based on ${\chi}_{\nu}^{2}$.\ \ The horizontal solid lines show the accepted values of $\tau$ and $q_0$ for percolation.}
\label{fig:12}
\end{figure}

Figure 13 shows the results of the single parameter fit analysis when applied to the same percolation system but with the cluster
multiplicity used as a measure of the control parameter.\ \ Again there is a minimum in the  ${\chi}_{\nu}^{2}$ values at some
intermediate value of the control parameter which indicates that $m_c = 57 \pm 3$ with $\tau = 2.2 \pm 0.1$, $q_0 = 0.20 \pm 0.01$
and a ${\chi}_{\nu}^{2} = 0.72$.\ \ Note the consistency between these values of $q_0$ and $\tau$ and those obtained with $p_l$
following this method.\ \ The location of the critical point determined here is consistent, at the 10$\%$ level, with the
insensitive signatures presented in the previous section.\ \ The lower ${\chi}_{\nu}^{2}$ value is due to the finer bins over which
the cluster distributions were grouped.\ \ Figure 16b shows the best fit power law.\ \  Here only clusters of size $A = 1$ and size
$A = A_{max}$ were excluded from the fitting procedure.

\begin{figure} [h]
\centerline{\psfig{file=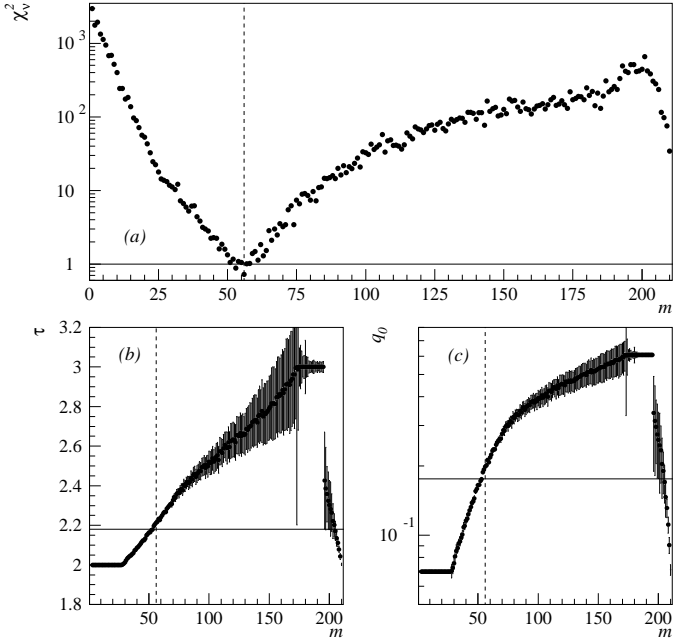,width=10.0cm,angle=0}}
\caption{Results for the full Fisher $\tau$-power law analysis on percolation as a function of $m$: (a) ${\chi}_{\nu}^{2}$,
(b) fitted ${\tau}$ and (c) $q_0$.\ \ The vertical dashed line shows the estimate of the critical point based on a best fit
based on ${\chi}_{\nu}^{2}$.\ \ The horizontal solid lines show the accepted values of $\tau$ and $q_0$ for percolation.}
\label{fig:13}
\end{figure}

From Figures 12 and 13 it could be argued, based on the best agreement between the fitted $\tau$ and the accepted three
dimensional percolation value, that there are better choices for the critical point than those quoted above.\ \ However, those
arguments assume knowledge of the value of $\tau$ as an input.\ \ The use of the location of the best fit to a single parameter
power law as an indicator of the critical point makes no assumption regarding the value of $\tau$ and is a test of the FDM
formalism in which only the range of $\tau$ is suggested: $2 < \tau < 3$.\ \ The values of $\tau$ and $q_0$ are outputs rather than
inputs of this analysis.\ \ Much of the following analysis presented in this paper follows the same philosophy.\ \ That is, the
analysis is designed to test the cluster distribution in question for behavior consistent with the FDM formalism.\ \ The values of
quantities, such as critical exponents, are results of the analysis method and are in no way selected for on the basis of their 
particular values.\ \ Agreement between exponent values determined by this procedure and the canonical values in various universality 
classes is then significant because the values of the exponents are determined solely by the behavior of the cluster distributions so 
analyzed.

The results of the single parameter power law fits for the random partitions are presented in Figure 14.\ \ There is a minimum in
the ${\chi}_{\nu}^{2}$ value at $m = 59$.\ \ However ${\chi}_{\nu}^{2} = 10.83$, which is an order of magnitude above the
percolation results, should not be used as an indication of a good fit of the cluster distribution by a single parameter power
law.\ \ The location of the ${\chi}_{\nu}^{2}$ minimum is also in disagreement with the insensitive signatures presented in the
last section.\ \ Here only clusters of size $A_f = 1$ and size $A_f = A_{max}$ were excluded from the fitting procedure.

\begin{figure} [h]
\centerline{\psfig{file=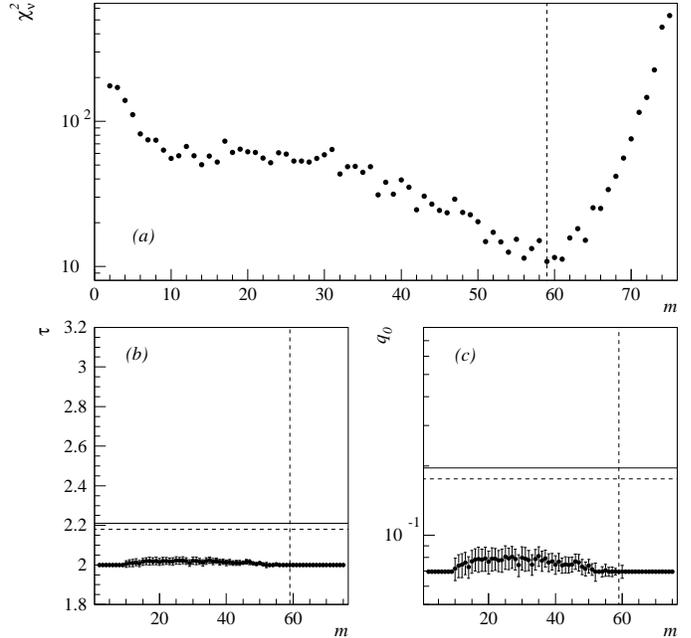,width=10.0cm,angle=0}}
\caption{Results for the full Fisher $\tau$-power law analysis on random partitions: (a) ${\chi}_{\nu}^{2}$, (b) fitted
${\tau}$ and (c) $q_0$.\ \ The horizontal solid (dashed) lines show the accepted values of $\tau$ and $q_0$ for percolation
(3D Ising).}
\label{fig:14}
\end{figure}

Figure 15 shows the results of this analysis applied to the gold multifragmentation data.\ \ As with the percolation results, the
${\chi}_{\nu}^{2}$ shows a minimum that drops nearly two orders of magnitude from the peaks for high and low $m$ to the valley at a
mid range value of $m$, see Figure 15a.\ \ In the context of the FDM analysis this result suggests that the critical point is
located at $m_c = 22 \pm 1$ with $\tau = 2.2 \pm 0.1$ and $q_o = 0.18 \pm 0.01$ and ${\chi}_{\nu}^{2} = 2.70$.\ \ The best fit
power law is show in Figure 16d.\ \ An uncertainty of one unit of multiplicity is assigned to $m_c$ to take into account the
relatively low values of ${\chi}_{\nu}^{2}$ of the neighboring fits. 

\begin{figure} [h]
\centerline{\psfig{file=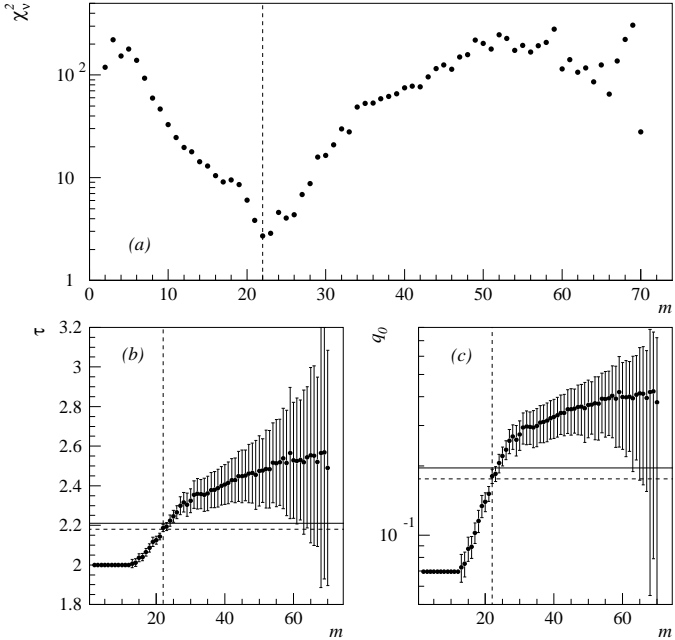,width=10.0cm,angle=0}}
\caption{Results for the full Fisher $\tau$-power law analysis on Au $+$ C multifragmentation: (a) ${\chi}_{\nu}^{2}$, (b)
fitted ${\tau}$ and (c) $q_0$.\ \ The vertical dashed line shows the estimate of the critical point based on a best fit based
on ${\chi}_{\nu}^{2}$.\ \ The horizontal solid (dashed) lines show the accepted values of $\tau$ and $q_0$ for percolation (3D
Ising).}
\label{fig:15}
\end{figure}

For the above fits to the gold multifragmentation data the ${\chi}_{\nu}^{2}$ is weighted by the errors in both $n_{A_{f}}$ and
$A_f$.\ \ The fitting procedure has also been performed with no error weighting on ${\chi}_{\nu}^{2}$ and with errors only in
$n_{A_{f}}$ for weighting.\ \ Both analyses shows results that were not significantly different from those quoted here.\ \  As
mentioned previously, clusters with $Z_f = 2$ are created in both the prompt first stage and in the multifragmenting of the gold
nuclear remnant.\ \ The prompt $Z_f = 2$ clusters have been excluded from the filtered gold multifragmentation analysis.\ \
However, as a further test of the single parameter power law fit, only clusters with $3 \le Z_f \le 16$, {\it i.e.} clusters with
no contamination from the prompt first stage, were included in a repeat of this analysis.\ \ Again the results show practically the
same behavior as those shown here.\ \ As yet another test, clusters with $2 \le Z_f < Z_{max}$ were included in the fitting
procedure, and again the results showed no difference from those presented here.\ \ Finally clusters with $3 \le Z_f < Z_{max}$
were included in the fitting procedure, and again the results showed no difference from those presented here.\ \ The data always
showed a deep valley in the ${\chi}_{\nu}^{2}$ versus $m$ plot which indicated that the location of the critical point was $m_c
\sim 22$ and that $\tau \sim 2.2$, $q_0 \sim 0.18$ and $1 < {\chi}_{\nu}^{2} < 4$.\ \ This analysis shows that the value of
${\tau}$ and the location of $m_c$ are not sensitive to the fit region.\ \ The behavior of the data show this clearly.\ \ See
open circles in Figure 16d.

\begin{figure} [h]
\centerline{\psfig{file=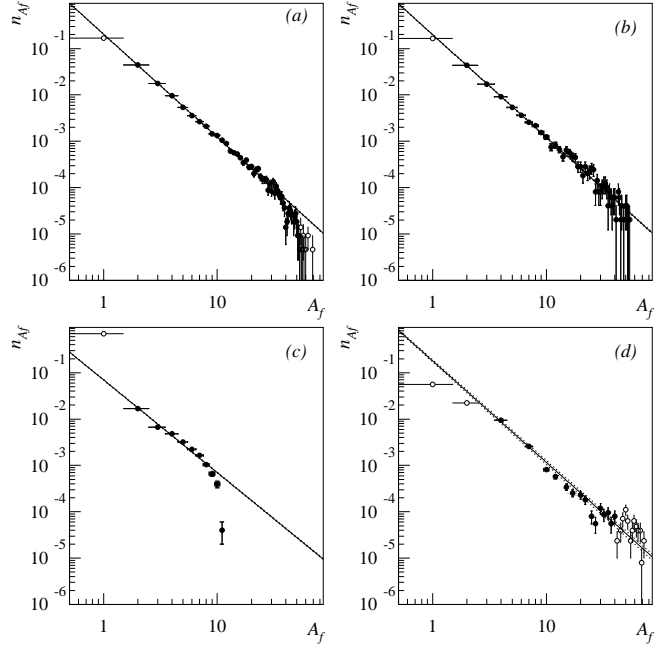,width=10.0cm,angle=0}}
\caption{Best fit Fisher $\tau$-power laws for: (a) percolation ($p_l$), (b) percolation ($m$), (c) random partitions and (d)
Au $+$ C multifragmentation.\ \ Solid circles were included in the fits, open circles were excluded.\ \ Solid lines show the best
fit power law, dotted lines indicate error in the power law based on errors in the fitted parameters.}
\label{fig:16}
\end{figure}

The single parameter power law analysis of the cluster distributions of the various systems produced the first result which can 
differentiate between systems that follow the FDM formalism and systems that do not.\ \ The differences between Figures 12a, 13a, 
15a and Figure 14a are clear and distinct.\ \ For both percolation and gold multifragmentation the behavior of
${\chi}_{\nu}^{2}$ is exactly what is predicted by the FDM formalism for continuous phase transitions.\ \ Far from the critical
point the cluster distribution is fit poorly by a single parameter power law due to the influence of a scaling function where
volume and surface effects overwhelm the underlying power law.\ \ At the critical point, where the influence of the scaling
function vanishes, the cluster distributions are well described by a single parameter power law with an exponent value, $\tau \sim
2.2$ and thus $q_0 \sim 0.2$, in keeping with what is expected for many universality classes.\ \ This fitting procedure does not
merely search out a cluster distribution which is well fit by a power law, but finds the cluster distribution which is well fit by
the FDM formalism.\ \ This is achieved via the coupling between the exponent $\tau$ and the normalization factor $q_0$.\ \ See eq.
(\ref{zeta_function}).\ \ The random partitions fail to produce such signals.\ \ This is expected as that system does not obey the
FDM formalism and thus should not show the same signals as systems that are known to follow the FDM such as percolation.\ \ This 
analysis of the cluster yield of gold multifragmentation yields a signal that is suggestive of critical phenomena.

\subsubsection{The critical exponent $\sigma$}

In section III it was shown that in the context of the FDM the surface of a cluster makes a contribution to the free energy of
cluster formation, via the scaling function $f(z)$, that depends on the number of constituents of the cluster raised to the power
$\sigma$.\ \ See eqn's (\ref{free_energy}), (\ref{surface_term_leath}) and (\ref{scaling_variable}).\ \ The behavior of the order 
parameter suggests that the scaling function $f(z)$ has a maximum \cite{stauffer_aharony}.\ \ At the maximum of the scaling function, 
$f_{max}(z_{max})$, the production of $A_f$ sized clusters is greatest:
     \begin{equation}
     n_{A_f}^{max}({\epsilon}_{max}) = q_0 A_{f}^{-\tau} f(z_{max}) .
     \label{sig_der_a1}
     \end{equation}
The argument of $f_{max}$ is:
     \begin{equation}
     z_{max} = A_{f}^{\sigma} {\epsilon}_{max},
     \label{sig_der_b}
     \end{equation}
where the value of $z_{max}$ depends on the specific details of the system in question \cite{stauffer_rep}.\ \ Rearranging eq.
(\ref{sig_der_b}) yields:
     \begin{equation}
     {\epsilon}_{max} = z_{max} A_{f}^{-\sigma}.
     \label{sig_der_c}
     \end{equation}
Thus ${\epsilon}_{max}$, the value of the control parameter at which the greatest number of clusters of size $A_f$ are produced, is
related to the cluster size through a simple power law with exponent $\sigma$.\ \ The exponent $\sigma$ can then be determined from
knowledge of the location of the critical point and the value of the control parameter at the greatest production of clusters of
size $A_f$.

The location of the critical point was determined in the search for the Fisher $\tau$-power law in section IV-B-1 and will be used
here to determine the $\sigma$.\ \ The value of the control parameter which yields the greatest production of each $A_f$ cluster
size was determined from the peak location in a plot of $n_{A_{f}}({\epsilon})$ versus the system's control parameter.\ \ See
Figure 17.

\begin{figure} [h]
\centerline{\psfig{file=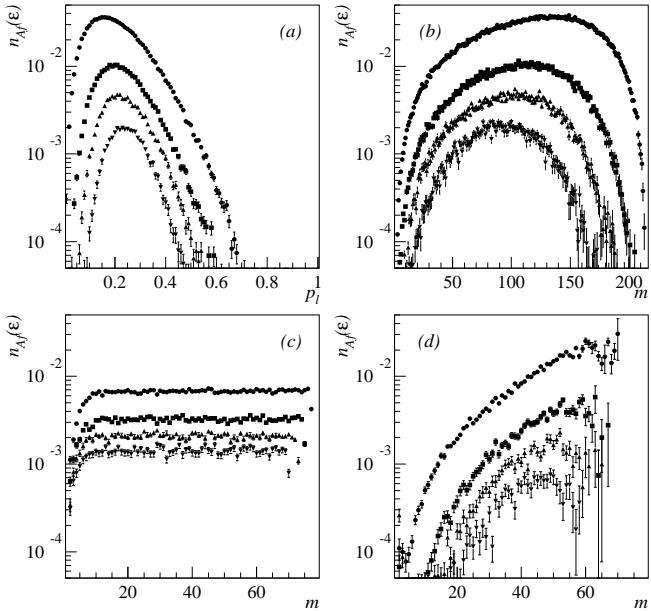,width=10.0cm,angle=0}}
\caption{Rise and fall in the production of individual clusters for: (a) percolation as a function of $p_l$, (b) percolation
as a function of $m$, (c) random partitions as a function of $m$ and (d) Au $+$ C multifragmentation as a function of $m$.\ \
Symbols (filled $\bigcirc$, $\Box$, $\bigtriangleup$, $\bigtriangledown$) indicate clusters of size (charge) 3, 5, 7 and 10 for
percolation and random partitions (multifragmentation).}
\label{fig:17}
\end{figure}

For each system at each cluster size plots such as those shown in Figure 17 were used to determine the location of the peak of
$n_{A_{f}}({\epsilon})$.\ \ For example, in percolation ($p_l$), ${\epsilon}_{max} = ( p_c - p_{max} ) / p_c$, pairs of points
$(n_{A_{f}}({\epsilon}), p_l )$ for a particular $A_{f}$ were fed into a SPLINE routine \cite{num_rec}.\ \ Input pairs were then 
{\it smeared} by assigning ${\delta}n_{A_{f}}({\epsilon})$ as the standard deviation of a gaussian centered on 
$n_{A_{f}}({\epsilon})$.\ \ Output of the SPLINE routine was used to interpolate the behavior of a smooth curve between the pairs 
of input points.\ \ Stepping along the interpolations in increments much smaller than the separation of the input $p_l$, a maximum 
of $n_{A_{f}}({\epsilon})$ was determined and $p_{max}$ was recorded.\ \ This process was repeated thousands of times for each cluster 
size and lead to an estimate of $p_{max} \pm {\delta}p_{max}$ as a function of $A_f$.

Using $p_{max}(A_f) \pm {\delta}p_{max}(A_f)$ and the value of $p_c \pm {\delta}p_c$, from the Fisher $\tau$-power law
determination process, the value of the exponent $\sigma$ was determined by taking the slope of $\ln({\epsilon}_{max})$
versus $\ln(A_f)$.\ \  The value of $z_{max}$ was determined by exponentiating the offset.\ \ The value of $p_c$ was varied
uniformly throughout the range suggested by ${\delta}p_c$ and tens of fits were made with varying starting and ending points in
$A_f$ of the fitting region.\ \  The final value of ${\sigma} \pm {\delta}{\sigma}$ and $z_{max} \pm {\delta}z_{max}$ are the
average and RMS values resulting from all the fits.

Results of this analysis performed on percolation ($p_l$) are shown in Figure 18a.\ \ Here the value of the control parameter that
coincides with the maximum in production of clusters of size $A_f$, ${\epsilon}_{max}$, is plotted against the cluster size.\ \ 
Results of the average power law fits to eq. (\ref{sig_der_c}) are plotted as a solid line.\ \ The agreement between the values 
returned by the procedure discussed above, ${\sigma} = 0.52 {\pm} 0.02$ and $z_{max} = -0.89 {\pm} 0.03$, and the accepted values 
for three dimensional percolation, ${\sigma} = 0.45$ and $z_{max} = -0.8$ \cite{stauffer_rep}, establishes the reliability of this 
exponent extraction method.\ \ The analysis differs in method from previous efforts on percolation lattices \cite{elliott_perc_1}, 
\cite{elliott_perc_2} but not in result.

\begin{figure} [h]
\centerline{\psfig{file=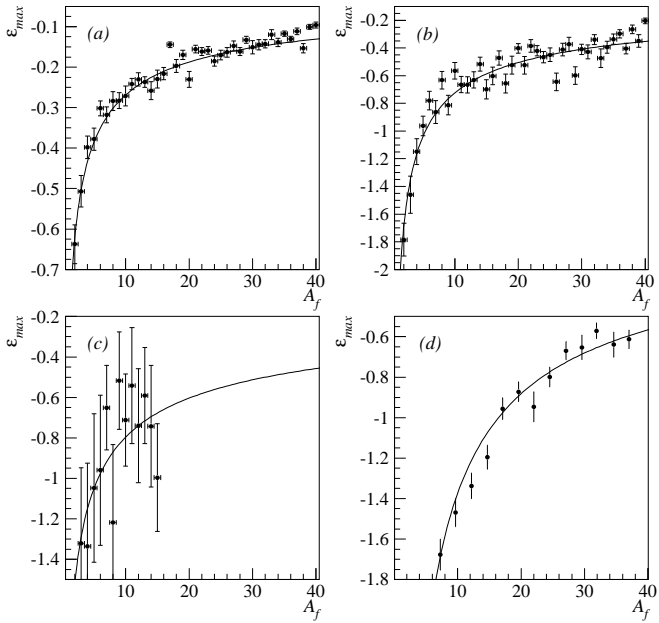,width=10.0cm,angle=0}}
\caption{Results of the $\sigma$ extraction procedure for: (a) percolation as a function of $p_l$, (b) percolation
as a function of $m$, (c) random partitions as a function of $m$ and (d) Au $+$ C multifragmentation as a function of $m$.}
\label{fig:18}
\end{figure}

The next test of this analysis is to extract a value of $\sigma$ from percolation ($m$).\ \ In order for this procedure to be
useful on multifragmentation data it must be shown that the exponent $\sigma$ can be determined using cluster multiplicity as the
control parameter.\ \ To that end the multiplicity at which the production of each cluster size is maximal, $m_{max}$ was
determined via the procedure described previously.\ \ Using the value of $m_c \pm \delta m_c$ determined via searching for the
Fisher $\tau$-power law and $m_{max}$ the exponent $\sigma$ was determined by taking the slope of $\ln({\epsilon}_{max})$ versus
$\ln(A_f)$.\ \ The value of $m_c$ was varied uniformly throughout the range suggested by ${\delta}m_c$ and several plots were made
with varying starting and ending points in $A_f$ of the fitting region.\ \ The value of $z_{max}$ was determined by exponentiating
the offset.\ \ Results of the average power law fits to eq. (\ref{sig_der_c}) are plotted as a solid line in Figure 18b.\ \ The
agreement between the values returned by this procedure, ${\sigma} = 0.52 {\pm} 0.02$, the value for $\sigma$ quoted in the above
paragraph and the accepted values for three dimensional percolation again establishes the reliability of this exponent extraction 
method and shows that the use of $m$ as a control parameter is acceptable.

The value of $z_{max} = -2.4 \pm 0.1$ extracted for percolation as a function of multiplicity is different from the value quoted
above, $z_{max} = -0.89 {\pm} 0.03$, for the percolation system as a function of probability.\ \ This is a result of changing the
measure of the control parameter from probability to multiplicity.\ \ This difference was observed in previous percolation efforts
\cite{elliott_perc_2} and explained therein.\ \ A plot of ${\epsilon}(p_{l})$ against ${\epsilon}(m)$ show that $z_{max}(p_{l})$
and $z_{max}(m)$ map to each other.\ \ See Figure 9 of ref. \cite{elliott_perc_2}.

Clusters from the random distribution were also subjected to this analysis.\ \ Due to the failure of the search for the Fisher
$\tau$-power law in the random partitions, the value of $m_c$ determined in the analysis of the gold multifragmentation was used,
$m_c = 22 \pm 1$.\ \ The value of the cluster multiplicity for maximum production of $A_f$ sized clusters was determined in the
same manner as with the percolation system.\ \ The flatness of the $n_{A_f}({\epsilon})$ versus $m$ curves, see Figure 17c, makes
finding a unique value of $m_{max}$ impossible.\ \ This is reflected in the large error bars on ${\epsilon}_{max} \pm \delta
{\epsilon}_{max}$ seen when plotted against $A_f$ in Figure 18c.\ \ The value of $m_{max}$ reported by the peak finding procedure
employed here reflects, approximately, the mid point of the multiplicity range of $n_{A_f}({\epsilon})$ for a particular $A_f$.\ \
Coupling the $m_c$ from the filtered gold multifragmentation data with the $m_{max}$ and fitting $\ln({\epsilon}_{max})$ versus
$\ln(A_f)$ gave $\sigma = 0.4 \pm 0.2$ and $z_{max} = -2.0 \pm 0.8$.\ \ However, it is clear when comparing the resulting average
fit for the random partitions shown in Figure 18c with either of the percolation results shown in Figure 18a and 18b that the
$\sigma$ resulting for the random partitions cluster distribution is meaningless.\ \ This is to be expected as the framework of the
FDM, used in the extraction of the exponent $\sigma$, is meaningful only when applied to systems which undergo a continuous phase 
transition.\ \  The {\it failure} of this analysis on this system is expected based on the basis of the {\it failure} of the analysis 
in the preceding section that aimed to find the Fisher $\tau$-power law and the critical point.

Results for the extraction of $\sigma$ from the gold multifragmentation data have been published previously \cite{elliott_sigma}, 
\cite{elliott_scaling}.\ \ In those analyses the largest cluster was excluded from consideration at every value of the control 
parameter.\ \ This is at odds with the formalism of the FDM where the sums excludes the largest cluster for $\epsilon > 0$ and include 
the largest cluster for $\epsilon < 0$. 

The previous analyses yielded values of $\sigma = 0.68 \pm 0.05$ and $0.65 \pm 0.06$ for work with the un-normalized charge 
distribution and normalized mass distribution respectively.\ \ When this analysis was redone using formalism of the FDM, {\it i.e.} 
the largest cluster excluded on one side of the critical point ({\it liquid}) and included on the other side ({\it gas}), the values 
of $\sigma$ were reduced by approximately $50$\%: $\sigma = 0.32 \pm 0.02$.\ \ In the case of percolation the difference introduced 
in the value of $\sigma$ when following the FDM formalism (as was done above) or not (as was the case in ref. \cite{elliott_perc_2}) 
is on the order of a few percent.\ \ This is the first notable difference observed in the qualitative behaviors of percolation cluster 
distributions and gold multifragmentation cluster distributions.

One source of this differing behavior is the changing mass of the system.\ \ For gold multifragmentation, from $A_0 \sim 194$ at low 
$m$ to $A_0 \sim 92$ at high $m$ \cite{hauger_prc}, while the system size is constant for percolation.\ \ For gold multifragmentation 
effects of the finite size of the system are felt more at high multiplicities than low.\ \ Since the percolation system size is 
constant, finite size effects are felt more evenly.

It is the higher values of $m$ where cluster production peaks in multifragmentation.\ \ The finite size of the system limits the size 
to which a cluster can grow.\ \ Thus the number of clusters of size $A_f$, $n_{A_f}$, as a function of $m$ is {\it contaminated} when 
the largest cluster, $A_{max}$, is included in a plot of $n_{A_f}$ versus $A_f$ because $A_{max}$ {\it would like} to be larger, but 
finite size effects limit the size $A_{max}$ can attain.\ \ Therefore, one method to account for this effect is to exclude $A_{max}$ 
from the cluster distribution at large $m$ values where these effects are largest\ \ This was done for the gold multifragmentation data.

The multiplicity at which the production of each cluster size is maximal, $m_{max}$ was determined via the procedure described 
previously.\ \ The value of $m_c$ determined in the Fisher $\tau$-power law analysis was used, $m_c = 22 \pm 1$.\ \ The value of 
$m_c$ was varied uniformly throughout the range suggested by ${\delta}m_c$ and several plots were made with varying starting and 
ending points in $A_f$ of the fitting region.\ \ The exponent $\sigma$ was determined by taking the slope of 
$\ln({\epsilon}_{max})$ versus $\ln(A_f)$ and the value of $z_{max}$ was determined by exponentiating the offset.\ \ The results 
were ${\sigma} = 0.64 {\pm} 0.05$ and $z_{max} = -6.0 {\pm} 0.8$, see Table II and III.\ \ The average power law fits are
shown in Figure 18d.

\begin{table*}
\begin{center}
\caption{Critical exponents}
\begin{tabular}{lllllll}
\hline
\multicolumn{1}{} 
.Exponent $/$ System       & 3D Percolation & Percolation ($p_l$) & Percolation ($m$) & Random Partitions & Au $+$ C & 3D Ising \\
\hline
$\tau$                    &  2.18 & $2.2  {\pm} 0.1 $  & $2.2  {\pm} 0.1 $ &  $2.0 {\pm} 0.1$  & $2.2  {\pm} 0.1 $ & 2.21 \\
$\sigma$                  &  0.45 & $0.52 {\pm} 0.02$  & $0.52 {\pm} 0.02$ &  $0.4 {\pm} 0.2$  & $0.64 {\pm} 0.05$ & 0.64 \\
$\gamma = {\frac{3-{\tau}}{\sigma}}$ & 1.82 & $1.5 {\pm} 0.2 $  & $1.5 {\pm} 0.2 $ &  $2.5 {\pm} 1.2$  & $1.3 {\pm} 0.2 $ & 1.23\\
$\gamma_+$ (matching)     &  & $1.8  {\pm} 0.2 $  & $1.64 {\pm} 0.04$ &  $0.4 {\pm} 0.1$  & $1.4  {\pm} 0.3 $ & \\
$\gamma_-$ (matching)     &  & $1.8  {\pm} 0.2 $  & $1.7  {\pm} 0.1 $ &  $0.5 {\pm} 0.2$  & $1.4  {\pm} 0.3 $ & \\
$\left<{\gamma}\right>$ (matching)   &  & $1.8  {\pm} 0.2 $  & $1.67 {\pm} 0.05$ &  $0.5 {\pm} 0.1$  & $1.4  {\pm} 0.3 $ & \\
$\Delta \gamma$ (matching)&  & $0.0  {\pm} 0.3 $  & $0.06 {\pm} 0.1 $ &  $0.1 {\pm} 0.2$  & $0.0  {\pm} 0.4 $ & \\
$\nu$ (hyperscaling)      &  0.87 & $0.77 {\pm} 0.07$  & $0.77 {\pm} 0.07$ &                   & $0.63 {\pm} 0.07$ & 0.63 \\
\hline
\end{tabular}
\end{center}
\end{table*}

\subsubsection{The scaling function $f(z)$}

With the critical point ($p_c$ or $m_c$), $\tau$, $q_0$ and $\sigma$ determined and assuming coexistence, $g = 1$, it is possible
to find the scaling function by rewriting eq. (\ref{cluster_distribution}) as
	\begin{equation}
	n_{A_{f}}(\epsilon) / q_{0} A_{f}^{-\tau} = f(z) .
	\label{scaling_isolate}
	\end{equation}
Doing this has the effect of appropriately scaling $n_{A_{f}}(\epsilon)$ and collapsing the data onto a single curve.\ \ Figure 19
shows the results of this sort of scaling.

In Figure 19a, b and d, the data from percolation ($p_l$ and $m$) and multifragmentation, respectively, show collapse onto a single 
curve for a wide range in cluster size and over nearly the full range in control parameter.\ \ Random partitions shows no such 
collapse, see Figure 19c.

As a demonstration of this type of scaling the same data has been scaled in the same fashion, but with a different choice of the 
critical point.\ \ Figure 20 shows the systems using a critical point with a value of half of the critical point determined via 
the Fisher $\tau$-power law, while Figure 21 shows the same analysis with a value of twice  the critical point determined via the 
Fisher $\tau$-power law.\ \ A visual inspection of Figures 19, 20 and 21 reveals the greatest data collapse occurs when the choice 
of the Fisher $\tau$-power law critical point is used, at least for the percolation ($p_l$ and $m$) and multifragmentation systems.\ \ 
Random partitions show no such collapse.\ \ Using different values of $\tau$ and $\sigma$ in this scaling analysis of random partitions 
does not significantly alter the data collapse.\ \ In Figures 19, 20 and 21 error bars on the data points are not shown for the sake of 
clarity.\ \ The size of the error bars reflect the scatter of the data and are larger for larger negative values of $z$ since there 
were lower statistics for higher multiplicity events \cite{hauger_prc}.

\begin{figure} [h]
\centerline{\psfig{file=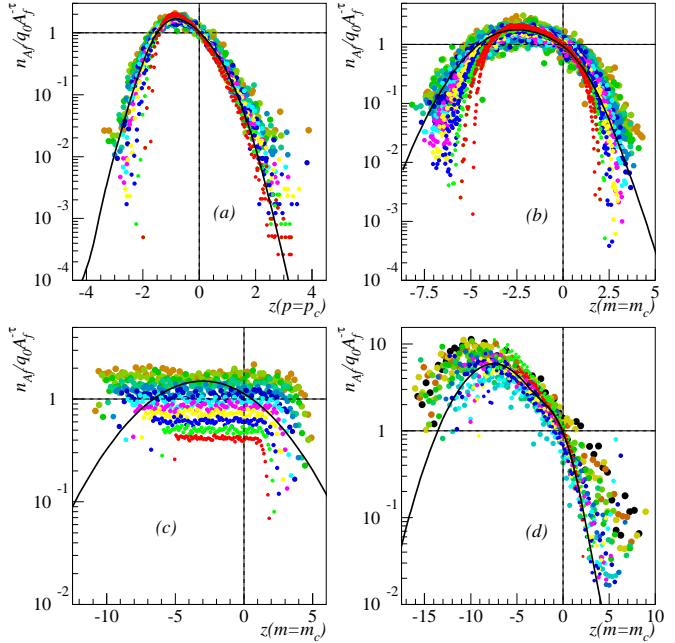,width=10.0cm,angle=0}}
\caption{Results of the scaling function analysis with the critical point equal to the value as determined from the Fisher
$\tau$-power law analysis for: (a) percolation as a function of $z(p_l)$, (b) percolation as a function of $z(m)$, (c) random
partitions as a function of $z(m)$ and (d) Au $+$ C multifragmentation as a function of $z(m)$.\ \ Solid curves show the fitted
scaling function.\ \ Smaller sized points show smaller sized clusters.\ \ Cluster sizes are most evident in (d) where at the bottom 
in red are $A_f = 2$ sized clusters and at the top in orange are $A_f = 14$.\ \ For percolation the colors cover a proportional range 
in cluster size while for multifragmentation the range is: red $Z_f = 3$ to orange $Z_f = 14$.}
\label{fig:19}
\end{figure}

\begin{figure} [h]
\centerline{\psfig{file=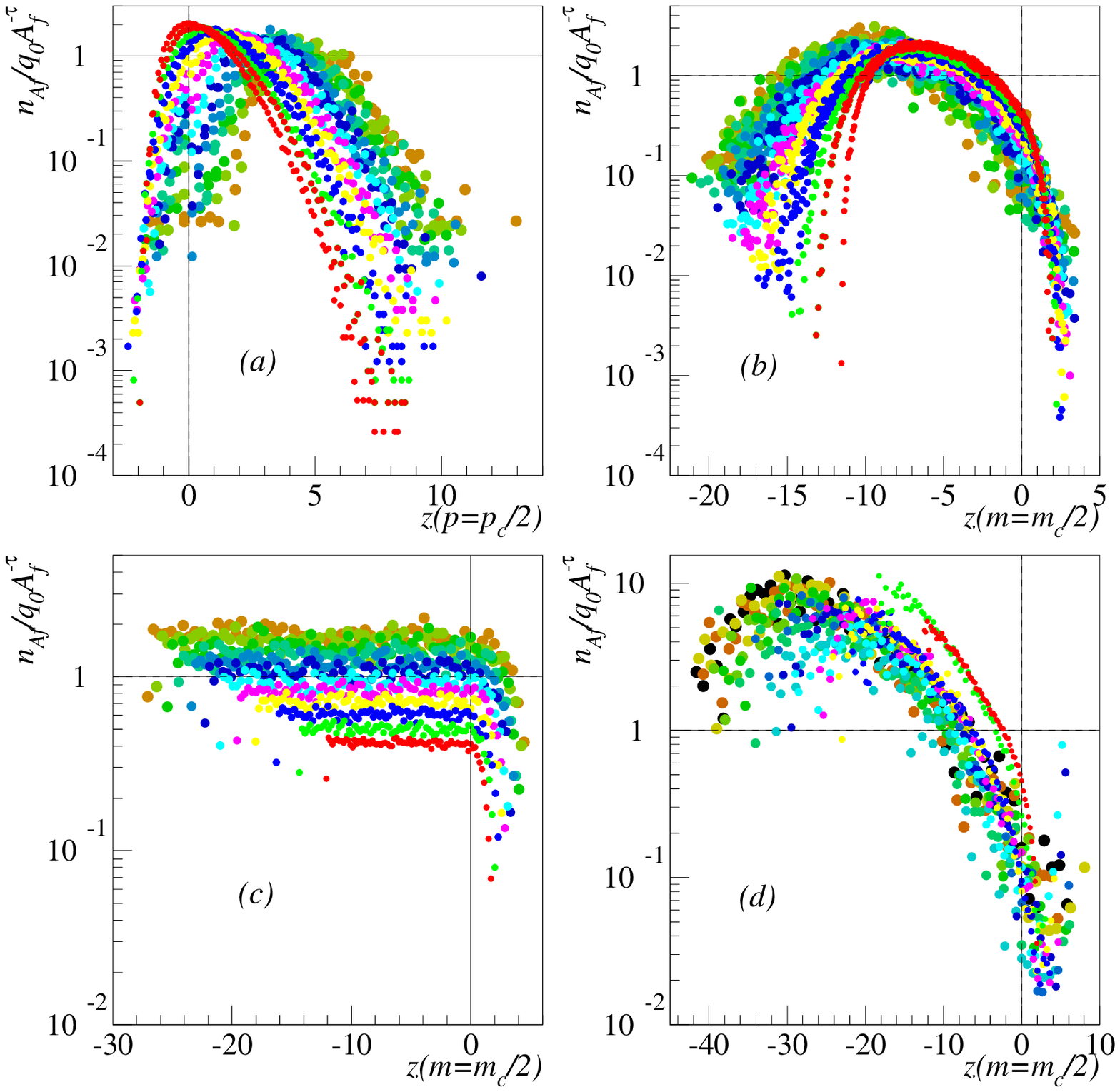,width=10.0cm,angle=0}}
\caption{Results of the scaling function analysis with the critical point equal to half the value as determined from the
Fisher $\tau$-power law analysis for: (a) percolation as a function of $z(p_l)$, (b) percolation as a function of $z(m)$, (c)
random partitions as a function of $z(m)$ and (d) Au $+$ C multifragmentation as a function of $z(m)$.}
\label{fig:20}
\end{figure}

\begin{figure} [h]
\centerline{\psfig{file=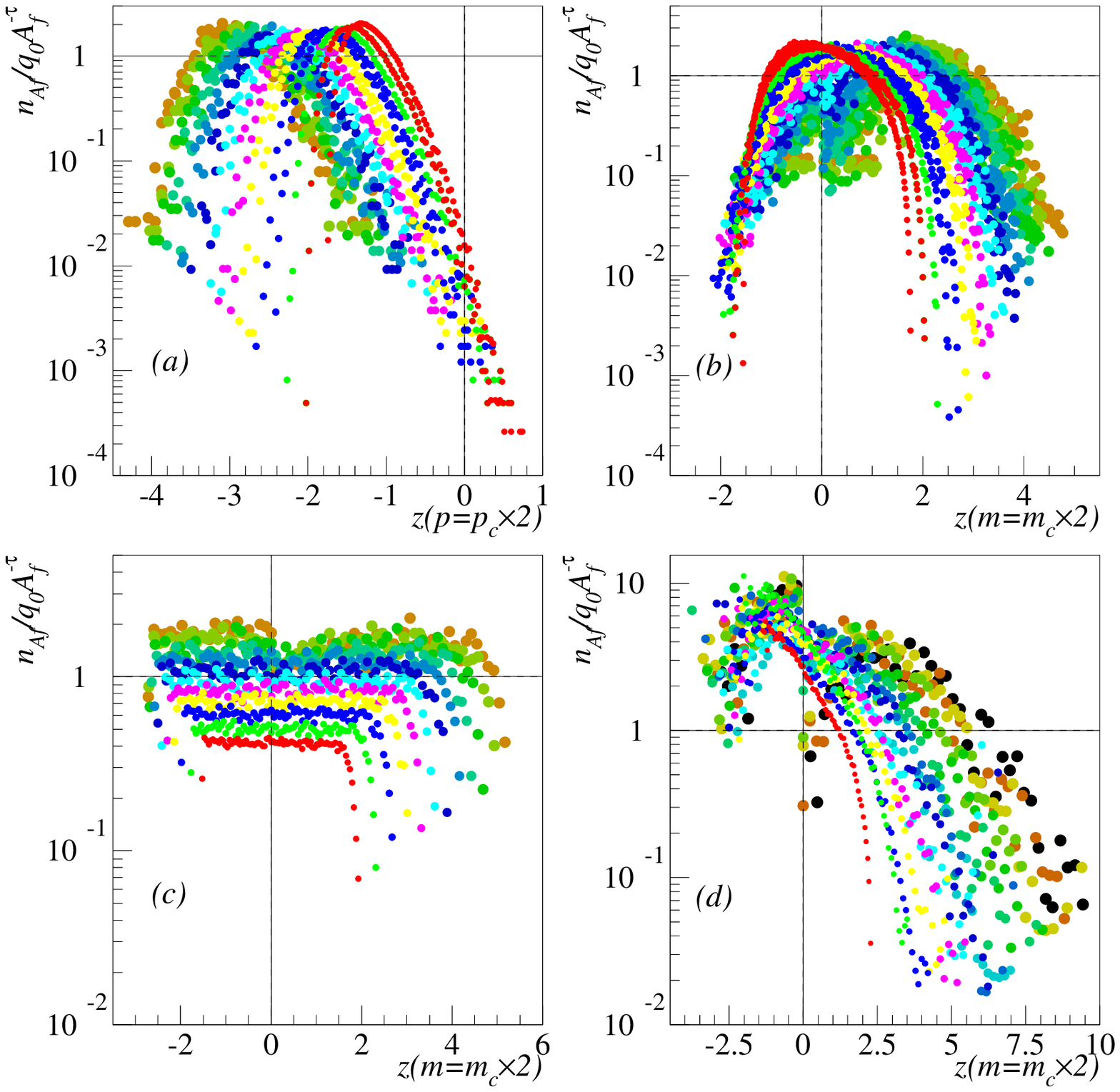,width=10.0cm,angle=0}}
\caption{Results of the scaling function analysis with the critical point equal to twice the value as determined from the
Fisher $\tau$-power law analysis for: (a) percolation as a function of $z(p_l)$, (b) percolation as a function of $z(m)$, (c)
random partitions as a function of $z(m)$ and (d) Au $+$ C multifragmentation as a function of $z(m)$.}
\label{fig:21}
\end{figure}

Figure 22 shows a quantitative measure of the data collapse from this scaling analysis.\ \ For a number of different choices of
control parameter scaling plots, as in Figures 19 through 21, were made.\ \ Each plot was binned along the abscissa and the RMS 
fluctuations for each bin were calculated.\ \ The RMS fluctuations in all bins were then summed and plotted as a function of the 
choice of critical point.\ \ See Figure 22.\ \ In the percolation ($p_l$ and $m$) and multifragmentation systems the data shows 
the most collapse in the neighborhood of the Fisher $\tau$-power law critical point.\ \ No such behavior is observed in the random 
partition system.\ \ This analysis serves as another, albeit crude, estimate of the location of the critical point.\ \ Table I
lists the results.

\begin{figure} [h]
\centerline{\psfig{file=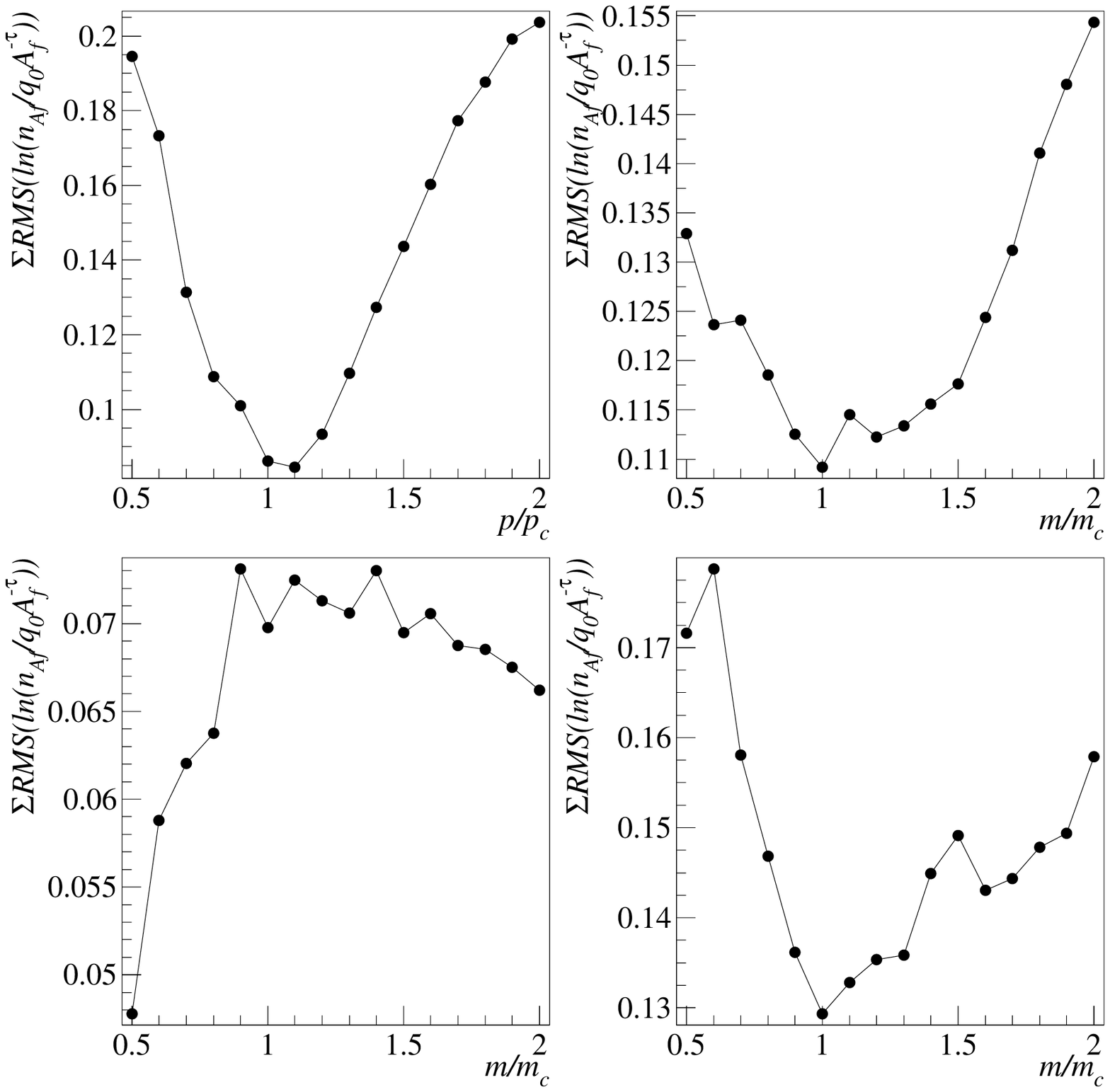,width=10.0cm,angle=0}}
\caption{A measure of the scatter in the scaling function analysis as a function of the choice of the critical point for: (a)
percolation ($p_l$), (b) percolation ($m$), (c) random partitions and (d) Au $+$ C multifragmentation.}
\label{fig:22}
\end{figure}

\begin{table*}
\begin{center}
\caption{Critical amplitudes}
\begin{tabular}{lllll}
\hline
\multicolumn{1}{}
.Amplitude $/$ System      &  Percolation ($p_l$)   &   Percolation ($m$)   &   Random Partitions    &  Au $+$ C   \\
\hline
$q_0$                     &   $ 0.20 {\pm} 0.01$   &   $ 0.20{\pm} 0.01$   &   $ 0.07 {\pm} 0.01$   &  $ 0.18 {\pm} 0.01$  \\
$z_{max}$                 &   $-0.89 {\pm} 0.03$   &   $-2.4 {\pm} 0.1 $   &   $-2.0  {\pm} 0.8 $   &  $-11.0 {\pm} 2.0 $  \\
$z_{max}$ (scaling fcn)   &   $-0.9  {\pm} 0.1 $   &   $-2.5 {\pm} 0.5 $   &   $-2.0  {\pm} 8.0 $   &  $-9.0  {\pm} 2.0 $  \\
${\Gamma}_{+}$ (scaling)  &   $ 0.9  {\pm} 0.1 $   &   $ 4.9 {\pm} 0.5 $   &   $ 3.5  {\pm} 0.5 $   &  $19.0  {\pm} 3.0 $  \\
${\Gamma}_{-}$ (scaling)  &   $ 0.17 {\pm} 0.05$   &   $ 0.3 {\pm} 0.1 $   &   $ 1.0  {\pm} 0.2 $   &  $ 0.24 {\pm} 0.05$  \\
${\Gamma}_{+}/{\Gamma}_{-}$ (scaling) & $5.0 {\pm} 2.0$  & $16.0 {\pm} 6.0$  & $3.5 {\pm} 0.9$      &  $80.0  {\pm}20.0 $  \\
${\Gamma}_{+}$ (matching) &   $ 1.0  {\pm} 0.3 $   &   $ 7.0 {\pm} 0.6 $   &   $26.0  {\pm} 2.0 $   &  $26.0  {\pm} 9.0 $  \\
${\Gamma}_{-}$ (matching) &   $ 0.08 {\pm} 0.07$   &   $ 0.28{\pm} 0.04$   &   $ 3.7  {\pm} 0.4 $   &  $ 0.27 {\pm} 0.06$  \\
${\Gamma}_{+}/{\Gamma}_{-}$ (matching) & $13.0 {\pm} 12.0$  & $25.0 {\pm} 4.0$  & $7.0 {\pm} 1.0$   &  $100.0 {\pm}40.0 $  \\
${\Gamma}_{+}$ (C.T.S. 3dP)&   $ 0.9  {\pm} 0.1 $   &   $ 6.4 {\pm} 0.5 $   &   $170.0 {\pm}20.0 $   &  $ 30.0 {\pm} 5.0 $  \\
${\Gamma}_{-}$ (C.T.S. 3dP)&   $ 0.06 {\pm} 0.01$   &   $ 0.38{\pm} 0.04$   &   $0.08  {\pm}0.01 $   &  $ 0.08 {\pm} 0.01$  \\
${\Gamma}_{+}/{\Gamma}_{-}$ (C.T.S. 3dP)& $15.0 {\pm} 3.0$  & $17.0 {\pm} 2.0$ &$2100.0{\pm}400.0$  &  $ 380.0{\pm}60.0 $  \\
${\Gamma}_{+}$ (C.T.S. 3dI)&                       &                       &   $140.0 {\pm}10.0 $   &  $ 55.0 {\pm} 5.0 $  \\
${\Gamma}_{-}$ (C.T.S. 3dP)&                       &                       &   $ 0.1  {\pm}0.01 $   &  $  0.28{\pm} 0.05$  \\
${\Gamma}_{+}/{\Gamma}_{-}$ (C.T.S. 3dI)&          &                       &   $1400.0{\pm}200.0$   &  $ 200.0{\pm}40.0 $  \\
${\Gamma}_{+}^{scaled}$ & & & &  $23.0  {\pm} 3.0 $  \\
${\Gamma}_{-}^{scaled}$  & & & &  $ 0.14 {\pm} 0.05$  \\
${\Gamma}_{+}^{scaled}/{\Gamma}_{-}^{scaled}$ & & & &  $170.0  {\pm}60.0 $  \\
\hline
\end{tabular}
\end{center}
\end{table*}

The scaled data were used to determine the functional form of the scaling function by fitting the data with an empirical
parameterization consisting of two gaussians instead of the single gaussian in eq (\ref{surface_term_leath}):  
\begin{eqnarray}
	f(z) & = & a_{1} \exp [- \frac{1}{2} (\frac{z-b_1}{c_1})^2] + \nonumber \\
             &   & a_{2} \exp [- \frac{1}{2} (\frac{z-b_2}{c_2})^2]. 
\label{scaling_fit}
\end{eqnarray} 
This was suggested by the asymmetry of the percolation ($p_l$) data, Figure 19a, and is consistent with a simplified version of
{\it corrections to scaling} \cite{margolina} as discussed in section V.\ \ Figure 19 shows the resulting fits for all systems.\ \
Fit parameter values can be found in Table IV.\ \ Errors on the parameters of the fits, {\it e.g.} $a_1$ etc., reflect the change
in those parameters when the range of clusters included was changed, {\it e.g.} clusters with $Z_f = 2$ were included or excluded
and so on, and the weighting on the fit was changed, {\it e.g.}  $\chi_{\nu}^2$ is unweighted, weighted with errors on
$n_{A_{f}}({\epsilon}) / q_{0} A_{f}^{-{\tau}}$ or with errors on $n_{A_{f}}({\epsilon}) / q_{0} A_{f}^{-{\tau}}$ and $\epsilon$.

\begin{table*}
\begin{center}
\caption{Scaling function parameters}
\begin{tabular}{llllll}
\hline
\multicolumn{1}{}
.Parameter $/$ System & Percolation ($p_l$) & Percolation ($m$) & Random Partitions &  Au $+$ C         &     Scaled Au $+$ C      \\
\hline                                                                               
$a_1$                & $ 0.8 {\pm} 0.2$    & $ 1.8 {\pm} 0.2 $ & $ 0.75{\pm} 0.5 $ &  $ 5.9 {\pm} 0.1$ &     $ 1.9 {\pm} 0.2$     \\
$b_1$                & $-1.0 {\pm} 0.1$    & $-2.6 {\pm} 0.3 $ & $-3.0 {\pm} 0.5 $ &  $-7.5 {\pm} 0.3$ &     $-3.1 {\pm} 0.3$     \\
$c_1$                & $ 0.5 {\pm} 0.2$    & $ 1.8 {\pm} 0.2 $ & $ 4.0 {\pm} 0.5 $ &  $ 3.2 {\pm} 0.1$ &     $ 1.8 {\pm} 0.2$     \\
$a_2$                & $ 1.0 {\pm} 0.1$    & $ 0.3 {\pm} 0.03$ & $ 0.75{\pm} 0.5 $ &  $ 0.8 {\pm} 0.2$ &     $ 6.9 {\pm} 0.7$     \\
$b_2$                & $-0.5 {\pm} 0.1$    & $ 0.1 {\pm} 0.01$ & $-3.0 {\pm} 0.5 $ &  $-1.2 {\pm} 0.4$ &     $-11.0{\pm} 1.0$     \\
$c_2$                & $ 0.8 {\pm} 0.2$    & $ 1.1 {\pm} 0.1 $ & $ 4.0 {\pm} 0.5 $ &  $ 1.5 {\pm} 0.2$ &     $ 3.8 {\pm} 0.4$     \\
\hline
\end{tabular}
\end{center}
\end{table*}

The scaling function for percolation ($p_l$ and $m$) determined here is {\it the} scaling function for percolation in three
dimensions, {\it i.e.} it is universal for three dimensional percolation independent of size.\ \ The scaling functions for $p_l$
and $m$ determined above agree well with the scaled cluster distributions of different size lattices, see Figure 23, and can be
used to predict the behavior of the second moment for any size lattice \cite{elliott_perc_2}.\ \ In the same spirit, the scaling
function determined here for gold multifragmentation is the scaling function for charged nuclear matter which describes the cluster
distributions produced in the multifragmentation of any nucleus, not just the excited gold remnant discussed in this work.\ \ With
the knowledge of the form of the scaling function various other quantities can be determined as illustrated in section III and
shown below.

\begin{figure} [h]
\centerline{\psfig{file=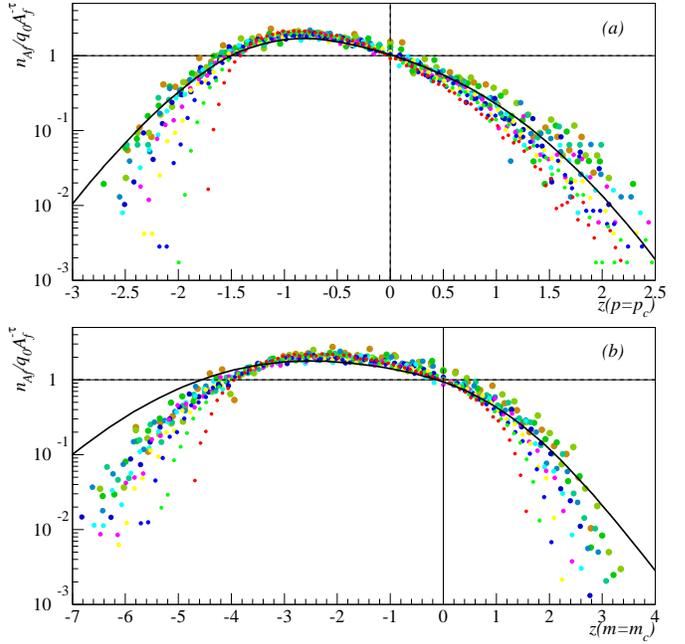,width=10.0cm,angle=0}}
\caption{The scaling function analysis for cluster distribution from a percolation lattice of $A_0 = 64$ as a function of (a)
$p_l$ and (b) $m$.\ \ Solid curves show the fitted scaling function from the $A_0 = 216$ lattice.}
\label{fig:23}
\end{figure}

The cluster distributions for the random partitions is fit, by eye, with the same empirical parameterization as in eq.
(\ref{scaling_fit}) see Figure 19c.\ \ The random partitions cannot be described by eq. (\ref{scaling_fit}).\ \ The solid curve 
in  Figure 19c will be used in the following section to demonstrate the failure of the scaling analysis, as is also seen here, 
when applied to a system where a continuous phase transition is absent.

Finally, a consistency check in this analysis is the agreement between the location of the peak in the scaling functions and the 
values of $z_{max}$ determined in the $\sigma$ analysis, see Table III.

\subsubsection{$\gamma$-power law from the scaling function}

The behavior of ${\kappa}_T$ or $M_2$ can be {\it derived} from the functional form of the scaling function and the critical
parameters via eq. (\ref{compressibility_f}).\ \ Performing the integration in eq. (\ref{compressibility_f}) using the functional
form of the scaling function determined above yields a direct calculation of the critical amplitudes, ${\Gamma}_{\pm}$ via eq.
(\ref{compressibility_g}).\ \ The critical exponent ${\gamma}$ is calculated from the values of ${\tau}$ and ${\sigma}$ via a
scaling relation in eq. (\ref{scaling_1}).\ \ Combining these two, ${\Gamma}_{\pm}$ and ${\gamma}$, it is possible to calculate
the ${\gamma}$-power law that describes the behavior of the second moment.\ \ This calculated ${\gamma}$-power law can then be
compared to the behavior of $M_2$ as measured from the cluster distribution.\ \ Figure 24a, b and d shows the agreement between 
the measured $M_2$ data (largest cluster omitted in the liquid region) and the calculated $\gamma$-power law curves for percolation 
($p_l$ and $m$) and gold multifragmentation, respectively and Tables II and III list the results.

The values of ${\gamma}$ determined via the scaling relation in eq. (\ref{scaling_1}) for percolation
($p_l$ and $m$) show approximate agreement with the accepted value of 1.8.\ \ The high value of $\sigma$ extracted above leads to a
low value of $\gamma$ here.\ \ Figures 24a and b also show the behavior of the second moment of a $250,047$ site lattice.\ \ The
power law predicted using the scaling function determined with a 216 site lattice shows rough agreement with the measured $M_2$ of
the larger lattice in both the amplitude ($\Gamma_{\pm}$) and exponent ($\gamma$).\ \ There is approximate agreement between the
predicted power law and the measured $M_2$ of the smaller lattice over some region in $\epsilon$ that is neither too near to, nor
too far from the critical point, ${\epsilon} = 0$.\ \ It is this region that will be determined, independently, in the following
section.

For the percolation ($p_l$ and $m$) system, the disagreement between the measured $M_2$ data and the calculated curves is due to two 
well known reasons: far from the critical point, the assumptions of scaling are no longer valid and the analytic background overwhelms 
the singular behavior.\ \ Near the critical point finite size effects dominate $M_2$, limiting the sizes of the large clusters which 
make the most significant contribution.\ \ In contrast, the ${\tau}$-power law was observed at the critical point because it is 
determined by smaller clusters which suffer the least from the finite size effects.

\begin{figure} [h]
\centerline{\psfig{file=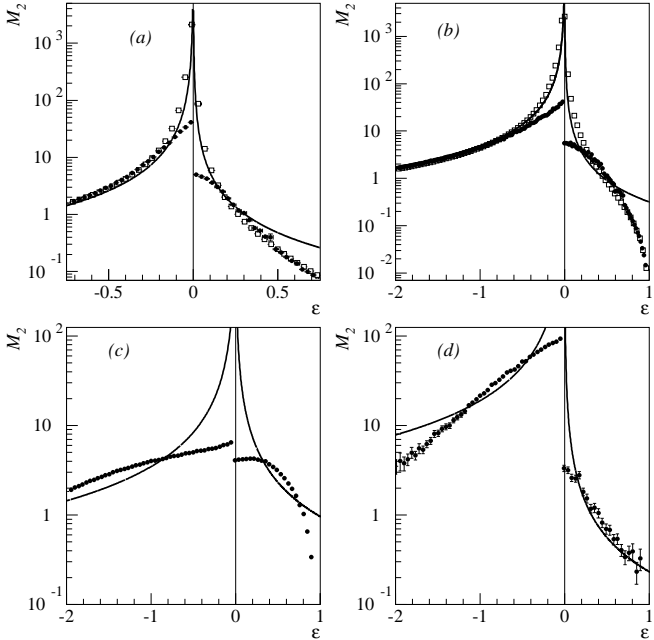,width=10.0cm,angle=0}}
\caption{Results of the $\gamma$-power law determined from the scaling function for: (a) percolation ($p_l$), (b) percolation
($m$), (c) random partitions and (d) Au $+$ C multifragmentation.\ \ Solid lines show the $\gamma$-power law predicted from the
scaling function, $\sigma$ and $\tau$.\ \ Filled circles show the second moment as a function of $\epsilon$.\ \ Open squares in (a)
and (b) show the second moment behavior of a percolation lattice of size $A_0 = 250,047$.}
\label{fig:24}
\end{figure}

Figure 24c shows the results when this analysis was applied to random partitions.\ \ The power law predicted from the scaling
function analysis applied to the cluster distribution of the random partitions fails to reflect the behavior of the measured second
moment.\ \ This is not surprising as the random partitions presented here are not the result of a system undergoing a continuous
phase transition.\ \ The disagreement observed in Figure 24c then serves as an indication of how this particular analysis probes
for the presence of a continuous phase transition.\ \ This figure shows the results of this analysis for a system with no phase 
transition, while Figures 24a and b show the results of this analysis on a system where such a phase transition is present.

The results of this analysis when applied to nuclear multifragmentation are shown in Figure 24d.\ \ In this case, the comparison to
the predicted $\gamma$-power law is neither as good as that for percolation nor as poor as that for the random partitions.\ \ It
shall be shown in section VI that considerable improvement can be achieved if account is taken of the changing system size, 
$A_{0}(m)$, and finite size scaling effects.

\ The approximate agreement between the predicted $\gamma$-power law and the measured $M_2$ behavior is in keeping with the
behavior expected for small systems undergoing a continuous phase transition, {\it e.g.} the percolation system.\ \ The
multifragmentation results are clearly different that then results of a system without a continuous phase transition, {\it e.g.}
random partitions.

\subsubsection{$\gamma$-matching}

In the previous works the procedure for determining critical exponent values and the location of the critical point from the
cluster distribution was based on a method of matching exponent values on both sides of the critical point \cite{gilkes_gamma},
\cite{elliott_perc_1}.\ \ The idea was to find the region on either side of the critical point where the power law behavior
predicted by the scaling function holds.\ \ As is seen in Figure 24 there is some intermediate $\epsilon$ region where the second
moment data are described by a power law, a region where the $M_2$ behavior is dominated by the $\gamma$-power law and all other
effects are small in comparison.\ \ In earlier percolation studies \cite{elliott_perc_1} general guidelines based on the 
correlation length and size of the fluctuations were used to find the boundaries in $\epsilon$ of the regions to be fit.\ \ In 
nuclear multifragmentation analyses \cite{gilkes_gamma} it was impossible to use such guidelines.\ \  Instead a method was 
developed that searched for regions best fit by power laws and determined the location of the critical point and exponent 
values simultaneously.\ \ The values of the critical exponents and the normalizations associated with power laws were obtained 
from the best fit power laws in those regions.\ \ As with the previous analyses presented in this paper, this method of exponent 
matching does not select a particular value of a critical exponent or the critical point.\ \ Instead the values found are the 
outcome of an unbiased procedure.

The method is as follows.\ \ A choice of the critical point, $p_c$ or $m_c$ was made.\ \ From this choice plots such as those shown
in Figure 24 were made.\ \  Then fitting boundaries in $\epsilon$ were chosen.\ \ The fitting range was defined by
$\epsilon^{far}_{\pm}$ and $\epsilon^{near}_{\pm}$.\ \ For example, on the gas side of the critical point a fit of $ln(M_{2})$
versus $ln(|{\epsilon}|)$ was made for all data with $|{\epsilon}^{near}_{+}| \le |{\epsilon}| \le |{\epsilon}^{far}_{+}|$.\ \
The slope of the resulting linear fit was recorded as ${\gamma}_{+}$, the offset as $ln({\Gamma}_{+})$ and the goodness of fit as
$\chi_{\nu +}^2$.\ \ The same procedure was applied to the liquid side of the chosen critical point, recording ${\gamma}_{-}$,
$ln({\Gamma}_{-})$ and $\chi_{\nu -}^2$.\ \ For each choice of the critical point, several choices of fitting regions,
$\epsilon^{far}_{\pm}$ and $\epsilon^{near}_{\pm}$ were made and results recorded.\ \ Five parameters were chosen for each set of 
power law regions examined: $\epsilon^{far}_{\pm}$, $\epsilon^{near}_{\pm}$ and $p_c$ or $m_c$.

The $\gamma$-power law fit regions and critical point locations were evaluated by demanding that: (1) they yield $\gamma_+$ and
$\gamma_-$  values that matched each other to within the error bars on those values returned by the fitting routine and (2) that the
$\chi_{\nu}^2$ of the fits were in the lowest quarter of the distribution resulting from all the fits which satisfy condition (1).\
\ The results from the power law fit regions that passed these two criteria were then histogrammed and average values for all
quantities concerned were determined.\ \ The results are summarized in Tables I, II and III and shown in Figure 25.

\begin{figure} [h]
\centerline{\psfig{file=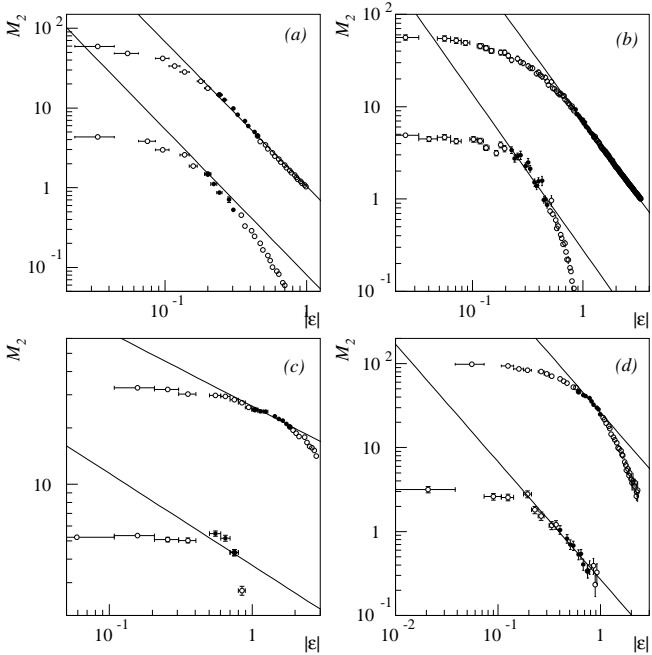,width=10.0cm,angle=0}}
\caption{Results of the $\gamma$-matching analysis for: (a) percolation ($p_l$), (b) percolation ($m$), (c) random partitions
and (d) Au $+$ C multifragmentation.\ \ Solid lines show the average $\gamma$-power fit returned by the procedure.\ \ Open circles
show the entire second moment behavior, filled circles show the average fitted region returned by the procedure.\ \ See text for
details.}
\label{fig:25}
\end{figure}

The lines plotted in Figure 25 do not result from any single fit, but display the average results for $\gamma_{\pm}$ and
$\Gamma_{\pm}$ that have satisfied conditions (1) and (2).\ \ The points in Figure 25 are the measured second moment for the
particular cluster distribution in questions plotted against $\epsilon$, which depends on the average value of $p_c$ or $m_c$ that
satisfies conditions (1) and (2).\ \ Therefore the lines in Figure 25 should not be interpreted as a fit to the data points shown
in the same figure, but as the average results from the $\gamma$-matching procedure.\ \ Full circles in Figure 25 show the average
fitting regions that satisfy conditions (1) and (2).

For percolation $p_l$ the value of $\gamma$ determined in this manner is within a few percent of the value determined in
\cite{elliott_perc_1} and the infinite lattice value.\ \ The ratio of $\Gamma_+ / \Gamma_-$ determined by this method, a ratio that
depends on the universality class of the system in question, is also in agreement with the infinite lattice value and the
$\Gamma_{\pm}$ values predicted by the scaling function, see Table III.\ \ The value of $p_c$ determined here is within 15$\%$ of 
the value determined in a previous analysis of the $L=6$ lattice \cite{elliott_perc_1} and the value determined above in the Fisher 
$\tau$-power law analysis, see Table II and Figure 25a.

The results for the analysis of percolation with $m$ as a measure of the control parameter are worse that the results when the 
natural control parameter $p_l$ is used, the difference in ${\gamma}_{+}$ and ${\gamma}_{-}$ was: ${\Delta}{\gamma}^{m} = 0.06 
\pm 0.1$ compared to ${\Delta}{\gamma}^{p_l} = 0.0 \pm 0.3$.\ \ This is to be expected because for each value of $p_l$ there is 
some spread in the resulting values of $m$, so that binning in $m$ groups together {\it events} with different values of $p_l$.\ \ 
There is also a non-linear relation between the average values of $p_l$ and $m$ \cite{elliott_perc_2}.\ \ In spite of these two 
effects the results of the analysis in section IV-B-4 suggests that vestiges of the signature of a phase transition are still present 
even when $m$ is used as the control parameter.\ \ That is also the case in the present analysis.\ \ Table II shows that the $\gamma$ 
value agrees, within error bars, with the infinite lattice value.\ \ The values of the critical amplitudes, $\Gamma_{\pm}$, do not 
yield a ratio that agrees with the infinite lattice value.\ \ This is due to the non-linear mapping of $p_l$ onto $m$ and is discussed 
in \cite{elliott_perc_2}.

When the $\gamma$-matching procedure was applied to the cluster distributions from random partitions a very limited amount of trial
fits passed the combined tests of (1) and (2).\ \ The results compared poorly to the percolation results.\ \ At best the
values of $\gamma_+$ and $\gamma_-$ match to within 20$\%$ of the average value of $\gamma$, compared to perfect matching for
percolation $p$ and matching within 5$\%$ for percolation $m$.\ \ The value of the critical point, $m_c$, returned from this
analysis also compared poorly to other outcome of previous analyses, see Table I.\ \ Finally, while fit regions for all systems were 
limited, the fit regions are the smallest and the poorest of quality for the random partitions.

The results of the $\gamma$-matching analysis applied to multifragmentation data has been published in ref. \cite{gilkes_gamma}.\ \
In that work the data were contaminated by the inclusion of prompt nucleons; prompt nucleons are excluded from consideration in
this work.\ \ In that work the second moment of the cluster distribution was determined based on the charge of a cluster rather than 
its mass as is done in this work.\ \ Previously, the second moment was generated from a cluster distribution that was not normalized 
to the changing size of the system as is done here.\ \ Furthermore the prior analysis consisted of only one quarter of the total number 
of events used in the present analysis.\ \ Thus the current analysis has higher statistics, has been freed of prompt nucleons, has a 
second moment that has been constructed with the masses from the cluster distribution and a cluster distribution that has been 
normalized to the changing system size.\ \ The exclusion of prompt nucleons and normalization to the changing system size are an 
effort to address the criticisms raised in \cite{bauer_3} and rebutted in \cite{gilkes_rep_1}.\ \ When the $\gamma$-matching procedure 
was applied to the data presented in this paper essentially the same results as presented in ref. \cite{gilkes_gamma} were recovered.\ \ 
See Table II and Figure 25d.\ \ One difference observed is in the value of the critical point returned, $m_{c}^{94} = 26 {\pm} 1$ 
reported in ref. \cite{gilkes_gamma} and $m_{c}^{99} = 21 {\pm} 2$ reported in this work.\ \ The difference is not as great 
as it appears to be.\ \ The origin of the published value of $m_{c}^{94}$ lies in picking the peak of the distribution of $m_c$ values 
that satisfied conditions (1) and (2) as the location of the critical point.\ \ The value was estimated based on the height of the peak 
and the error based on the width of the peak.\ \ The mean and RMS of the $m_c$ distribution in ref. \cite{gilkes_gamma} 
suggest a value of the critical point of $m^{94}_{c} = 25 {\pm} 3$.\ \ This value agrees, to within error bars, with the value 
of $m^{99}_{c}$ presented here.\ \ The relatively small shift in $m_c$ can then be understood to arise from the differences in the data 
sets.\ \ Noting this it is clear that the present $\gamma$-matching analysis is in agreement with the previous work.

The results of the present work are, again, in keeping with the expected results of a small system undergoing a continuous phase
transition.\ \ There is some region where matching $\gamma$ values can be obtained, some regions in ${\epsilon}$ where th
${\gamma}$-power law overwhelms all other effects.\ \  The fits in Figure 25d are of quality than those for random
partitions in Figure 25c and cover a greater range.\ \ When compared to the percolation $m$ results the multifragmentation data
compare favorably in terms of overall goodness of fits, width of fit region and matching of $\gamma_{\pm}$.\ \ See Table II.\ \ The
location of the critical point returned by this analysis also compares well with the location from other analyses.\ \ See Table I.

\section{Corrections to Scaling}

In the last section it was seen that the $\gamma$-power law and the data for the second moment in all systems have agreed over only
a limited area.\ \ To some degree this is to be expected.\ \ Near the critical point, assumptions valid for thermodynamic systems
are invalid for the finite systems discussed in this work.\ \ For that reason, finite size effects dominate at the critical point
and the second moment merely peaks instead of diverging.\ \ Far from the critical point other effects come into play.\ \ The
scaling assumptions inherent in the FDM are valid only in the neighborhood of the critical point.\ \ The size of this neighborhood
is somewhat ill defined and seems to depend on many factors, {\it e.g.} the quantity in question, the nature of the system, the
size of the system and so on.\ \ Scaling behavior in physical systems can be observed over a wide range in temperatures and
densities.\ \ This is most elegantly illustrated in the Guggenheim Plot \cite{guggenheim} of scaled temperature ($T/T_c$) as
function of scaled density (${\rho}/{\rho}_c$) for several different gases (Ne, Ar, Kr, Xe, N$_2$, O$_2$, CO and CH$_4$).\ \ In
that plot the data collapse onto a curve that is well described by a power law with an exponent of $\beta = 1/3$.\ \ The range in
validity of this agreement between data and power law is shown on the Guggenheim Plot to be over a range of $\Delta T \sim 0.5 T_c$
and $\Delta \rho \sim 2.5 \rho_c$.\ \ However, another system, the combination of isobutyric acid and water, shows the Guggenheim
type of scaling only very near the critical point \cite{greer}.\ \ Already when the range considered is $\Delta T \sim 0.04 T_c$
and $\Delta \rho \sim 0.01 \rho_c$ corrections to scaling can be observed.\ \ To that end, higher order corrections to scaling are
now examined in order to determine if fits such as those shown in previous sections can be improved.\ \ However, any improvement
comes at the expense of more fit parameters and assumptions.

To fully explore corrections to scaling in the context of the present systems where the cluster distributions serve as the main
observable the FDM is revisited in a fashion employed in references \cite{margolina} and \cite{adler}.\ \ Assuming coexistence eq.
(\ref{cluster_distribution}) is then re-written as
     \begin{equation}
     n_{A_{f}}({\epsilon}) = q_{0} A_{f}^{-{\tau}} \left ( f_{0}(z) + A_{f}^{-\Omega} f_{1}(z) + \ldots \right ) ,
     \label{corrected_cluster_distribution}
     \end{equation}
where $f_{1}(z)$ is the correction-to-scaling function and $\Omega$ is the correction-to-scaling exponent.\ \ The form of eq.
(\ref{corrected_cluster_distribution}) anticipates the presence of a second function of $z$.\ \ In section IV-3 it was found
empirically that both the scaled percolation and multifragmentation cluster distributions ($n_{A_f}({\epsilon}) / q_0
A_{f}^{-\tau}$) could be reasonably well described by the sum of two gaussians, eq. (\ref{scaling_fit}).\ \ In that treatment, the
amplitude of each gaussian was a constant, $a_1$ and $a_2$.\ \ If $A_f$ is restricted to a single value, the prescription give by
eq. (\ref{corrected_cluster_distribution}) is equivalent to that of eq. (\ref{scaling_fit}).\ \ Eq.
(\ref{corrected_cluster_distribution}) predicts that there should be an ordering to the scaled cluster distributions, {\it i.e.}
smaller cluster sizes should lie above the larger clusters due to the correction term.\ \ This can be observed in Figures 19a and b
in the neighborhood of the maximum of the scaled data for the percolation systems.\ \ In the tails of the distribution, either
large or small cluster production is suppressed.\ \ In the case of multifragmentation data, Figure 19d, the ordering is generally
observed where the statistics are adequate, namely, near $z = 0$.\ \ The ordering of the random partitions implies that $\Omega <
0$.

From eq. (\ref{corrected_cluster_distribution}) it possible to derive the {\it corrected} isothermal compressibility (second
moment) power law.\ \ Following the method in section III leads to:
     \begin{eqnarray}
     {\kappa}_{T} & \sim & ( {\rho}^2 k_{b} T )^{-1}  \times \nonumber \\
     &   & \left( \left | {\frac{q_0}{\sigma}} \int_0^{\pm \infty} dz \:f_{0}(z) \:|z|^{\frac{3 - \tau - \sigma}{\sigma}} \right | \right. + \nonumber \\
     &   & \left. \left | {\frac{q_0}{\sigma}} \int_0^{\pm \infty} dz \:f_{1}(z) \:|z|^{\frac{3 - \tau - \sigma - \Omega}{\sigma}} \right | 
           |\epsilon|^{\frac{\Omega}{\sigma}} \right) \nonumber \\
     &   & \times |\epsilon|^{\frac{\tau - 3}{\sigma}}
     \label{corrected_compressibility_1}
     \end{eqnarray}
which is usually simplified and written as
     \begin{equation}
     {\kappa}_{T}  \sim \Gamma_{\pm} |\epsilon|^{-\gamma} \left ( 1 + a_{\pm} |\epsilon|^{\Delta} \right ) .
     \label{corrected_compressibility_2}
     \end{equation}
Now the overall amplitude, $\Gamma_{\pm}$ is given by the first integral, and the correction-to-scaling amplitude is given by the
second integral divided by the first.\ \ The correction-to-scaling exponent is $\Delta = \Omega / \sigma$.

Using eq. (\ref{corrected_compressibility_2}) to fit the second moment distribution would lead to determining four fit parameters:
two amplitudes and two exponents.\ \ To explore the effects of corrections to scaling an assumption was made as to the universality
class of the system in question and thus the choice of exponent values.\ \ For the three dimensional percolation universality class
$\Delta = 1.22$ \cite{margolina}, \cite{adler} and for the three dimensional Ising universality class $\Delta = 0.56$
\cite{saul}-\cite{pestak}.\ \ The amplitudes were left as free parameters and the second moment of the cluster distributions were
fit.\ \ The value of the critical point determined from the $\gamma$-matching analysis of section IV-B-5 was used for each system.\
\ Figure 26 shows the results.

\begin{figure} [h]
\centerline{\psfig{file=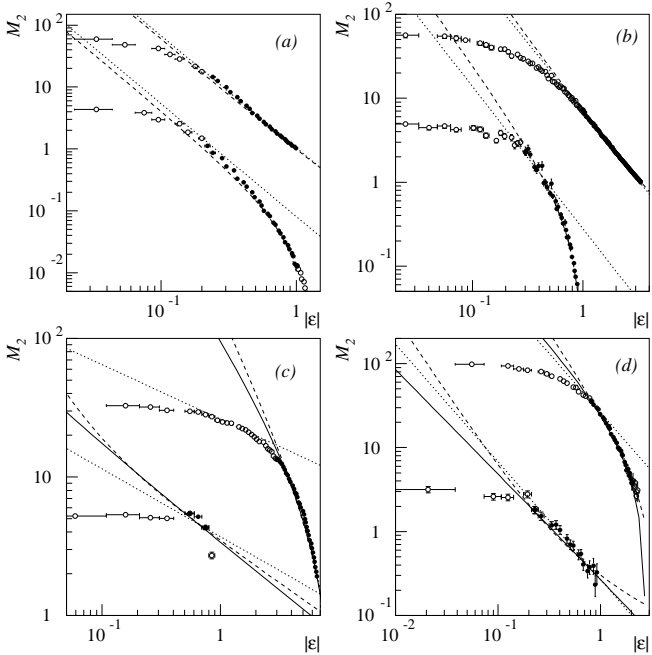,width=10.0cm,angle=0}}
\caption{Results of the correction-to-scaling analysis for: (a) percolation ($p_l$), (b) percolation ($m$), (c) random
partitions and (d) Au $+$ C multifragmentation.\ \ Open circles show the entire second moment behavior, filled circles show the
fitted region used in this procedure.\ \ See text for details.\ \ Dotted lines show the $\gamma$-matching power law.\ \ Dashed
(solid) lines show the resulting correction-to-scaling $\gamma$-power law for the 3D percolation (3D Ising) universality class.}
\label{fig:26}
\end{figure}

For the percolation system with the corrections-to-scaling a better fit to the second moment data was possible over a greater range
in $\epsilon$, up to twice the range of the average fitted region in the $\gamma$-matching analysis, see Figures 26a and b.\ \ The
fits still failed to reproduce the behavior of $M_2$ near the critical point where finite size effects dominate the system.\ \
Table III lists the results for the critical amplitudes, $\Gamma_{\pm}$.\ \ The agreement in the critical amplitudes determined in
this analysis and the amplitudes from the $\gamma$-matching analysis is due to the agreement of the behavior of eq.
(\ref{corrected_compressibility_2}) and the $\gamma$-power law from the $\gamma$-matching analysis over the region in $\epsilon$
determined by $\gamma$-matching.\ \ Thus the $\gamma$-matching analysis finds regions that are the least affected by higher-order 
corrections to scaling.

For the random partitions an improvement is only observed for the high multiplicity region where a better fit over a larger range
was obtained for both choices of universality class, see Figure 26c.\ \ The low multiplicity events showed no such improvement
partly due to the limited range in $\epsilon$ available.\ \ Both the three dimensional percolation and three dimensional Ising
exponents were used in this analysis for the random partitions and the multifragmentation data.\ \ Both choices of universality
classes showed similar results.\ \  The lack of effect of corrections to scaling is to be expected in a system that does not follow
FDM like scaling laws.

The multifragmentation data also showed improvement resulting in a better agreement between the fits and the $M_2$ data points over
a larger range in $\epsilon$, see Figure 26d.\ \ The improvement was observed for both choices of universality classes thus
indicating this analysis is insensitive to the differences \cite{bauer_4}, \cite{elliott_rep_1}, though the goodness of fit was
better for the choice of the three dimensional Ising exponents over the same fit regions (3D Ising: $\chi_{\nu+}^2 = 0.4$ and
$\chi_{\nu-}^2 = 0.7$; 3D percolation: $\chi_{\nu+}^2 = 0.7$ and $\chi_{\nu-}^2 = 1.1$).\ \ At this level of analysis it appears
that corrections to scaling improves the fits for the $\gamma$-power law.\ \ Whether this is due to the presence of a continuous
phase transition in nuclear multifragmentation, or merely extra terms in the fitting function remains an open question.

\section{Changing System Size and Finite Size Corrections}

It has been pointed out that the previous statistical analysis of gold multifragmentation \cite{gilkes_gamma} ignored the changing
size of the system \cite{bauer_3}.\ \ To first order this may have been a reasonable procedure \cite{gilkes_rep_1} as many
statistical signatures of a continuous phase transition have been observed both before and after the scaling to account for the
changing system size has been performed; {\it e.g.} the $\gamma$ power law shown here and in previous works agree well.\ \ However,
the data collapse in Figure 19d is qualitatively not as great as that shown by the percolation ($m$) system in Figure 19b and the
agreement between the calculated and measured $M_2$ behavior in gold multifragmentation in Figure 24d is qualitatively not as good
as that shown by percolation ($m$) in Figure 24b.\ \ In this section, the effect of the changing size of the system is explored and
accounted for.

The size of the multifragmenting system is shown in Figure 27a after ref. \cite{hauger_prc}.\ \ An approximately linear relation
between the system size, $A_0$, and $m$ was found: $A_0 = 199 - 1.6m$, see Figure 27a.\ \ The functional form of $A_{0}(m)$ was
used in the following analysis to account for the changing system size in an average way, {\it i.e.} not on an event-to-event basis.

If the multifragmenting system is assumed to be a system undergoing a phase transition, then the theory of finite size scaling of
the critical point \cite{fisher_a}-\cite{barber} asserts that the effective critical temperature, $T_c (A_0 )$, changes as a
function of the system size.\ \ Coupling this with the changing size of the system indicates that at each value of $m$ the value of
$T_{c}^{eff}(A_0)$ is different.

\begin{figure} [h]
\centerline{\psfig{file=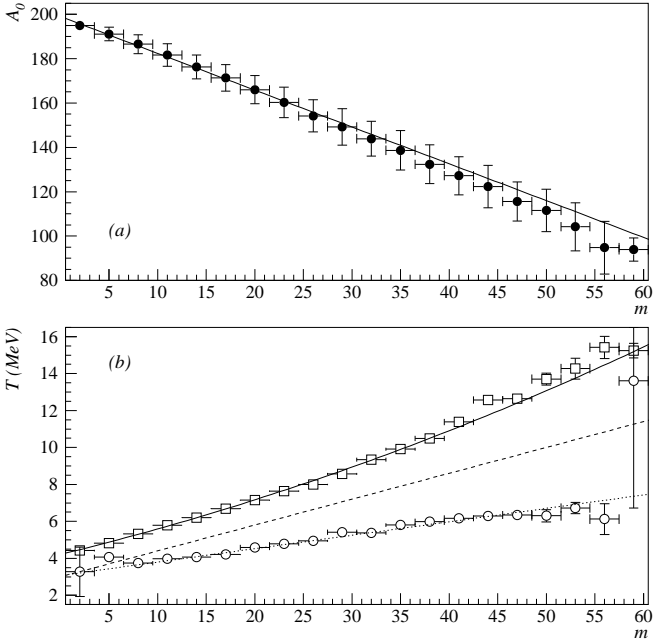,width=10.0cm,angle=0}}
\caption{(a) Behavior of the size of the system, $A_0$, with respect to the total charged particle
multiplicity, $m$.  The solid line is a fit to the changing system size.\ \ (b) measures of the temperature, $T$, as a function of
$m$: $\Box$ $T_i$ from a Fermi gas and $\bigcirc$ $T_f$ from an isotope thermometer.\ \ The solid curve shows a fit to $T_i$,
dotted curve a fit to $T_f$ and the dashed curve a fit to $ \left< T \right> = ( T_i + T_f ) / 2 $. }
\label{fig:27}
\end{figure}

The value of $T_{c}^{eff}(A_0)$ can be determined in the following manner.\ \ First a relation between the multifragmenting
system's temperature, $T$, and $m$ must be determined.\ \ Again from ref. \cite{hauger_prc} a relation can be found, see Figure
27b.\ \ Two estimates of $T$ were made, one from a Fermi gas (uncorrected for the effect of expansion energy), $T_i$, and the
other from an isotopic yield ratio thermometer, $T_f$, \cite{albergo}.\ \ These temperatures give an approximate indication of the
initial and final temperatures of the system.\ \ Figure 27b shows that $T_i$ is well fit by a quadratic function, while $T_f$ is
well fit by a linear function.\ \ Another linear function reproduces the average of $T_i$ and $T_f$: $T = 3.0 + 0.14 m$, this
was used for the following analysis.\ \ The critical temperature of infinite nuclear matter was assumed to be $T_{c}^{\infty} =
17 \pm 1$ MeV \cite{palmer_anderson}.

From the Fisher $\tau$-power law analysis the value of the multiplicity at the critical point was determined to be $m = m_c = 22
\pm 1$.\ \ The system size at that point is then $A_0 (m=22) = 164 \pm 2$ and the temperature is $T = 6 \pm 2$ MeV.\ \ This indicates 
that the critical temperature for a nuclear system with 164 nucleons, $T_{c}^{eff}(A_0)$, is approximately $6 \pm 2$ MeV.

According to theory, to first order the critical point scales with system size as:
     \begin{equation}
     ({T}_{c}^{eff}(A_0) - T_{c}^{\infty} ) / T_{c}^{\infty} = b {A}_{0}^{-1/d{\nu}},
     \label{fss}
     \end{equation}
where $d$ is the Euclidian dimension of the system and $\nu$ is the so-called hyperscaling exponent.\ \ At the smallest of system
sizes higher order correction terms may play an important role in the scaling of the critical point \cite{ferrenberg}.\ \ It is
assumed in this procedure that the above mentioned finite size scaling dominates all other effects, including those which arise due
to the charge of the protons present in the excited remnant.\ \ This is in keeping with the general philosophy of the theory of
critical phenomena where near the critical point the precise details of the system are irrelevant.

If it is assumed that gold multifragmentation is the result of a continuous phase transition and that transition occurs in three
dimensions, $d = 3$, then using the {\it hyperscaling relation} \cite{stauffer_aharony}
     \begin{equation}
     {\nu} = {\frac{{\tau} - 1}{d {\sigma}}},
     \label{hyperscaling}
     \end{equation}
with the extracted values of $\sigma$ and $\tau$, gives $\nu = 0.63 {\pm} 0.07$.\ \ The coefficient $b$ in eq. (\ref{fss}) can be
determined using $T_{c}^{eff}(A_0(m_c))$, $T_{c}^{\infty}$, $d$ and $\nu$; resulting in $b = -9 \pm 2$.\ \ Note that this value of 
$b$ suggests that $T_{c}^{eff}(A_0) = 0$ for $A_0 = 70 \pm 40$.\ \ This is a result of the form of eq. (\ref{fss}) and the notion 
that the critical temperature lowers as the size of the system decreases.\ \ Presumably higher order effects not taken into account 
in eq. (\ref{fss}) will affect the location in system size, $A_0$, where the effective critical temperature vanishes.\ \ For the 
form of finite size scaling corrections shown in eq. (\ref{fss}), only $b = -1$ yields an effective critical temperature that vanishes 
at $A_0 = 1$.

Now eq. (\ref{fss}) can be used to solve for $T_{c}^{eff}(A_0(m))$ and thus determine the scaled control parameter:
     \begin{equation}
     {\epsilon}^{scaled} = ({T}_{c}^{eff}(A_0(m)) - T ) / {T}_{c}^{eff}(A_0(m)).
     \label{fss_cp}
     \end{equation}

The analysis to extract the exponent $\sigma$ was performed with this ${\epsilon}^{scaled}$ by finding the peak in $A_f$ sized
cluster production as a function of $n_{A_f}({\epsilon}^{scaled})$ versus ${\epsilon}^{scaled}$.\ \ Previously, it was argued 
that the largest cluster should be excluded for all values of $\epsilon$ to account for finite size effects in the analysis to 
extract $\sigma$.\ \ The analysis in this section directly accounts for finite size effects, thus the standard FDM formalism 
with respect to the largest cluster is followed.\ \ This results in a value of $\sigma = 0.65 \pm 0.07$ and $z_{max}^{scaled} 
= -11 \pm 2$, Figure 28a shows the resulting power law. 

The scaling function for gold multifragmentation was then plotted using the above corrections for the changing system size and
finite size scaling, see Figure 28b.\ \ The data collapse is qualitatively better than in Figure 19d.\ \ The two gaussian
parameterization of $f(z^{scaled})$ was fit to the scaled scaling function and is shown in Figure 28b with parameters listed in
Table IV.

\begin{figure} [h]
\centerline{\psfig{file=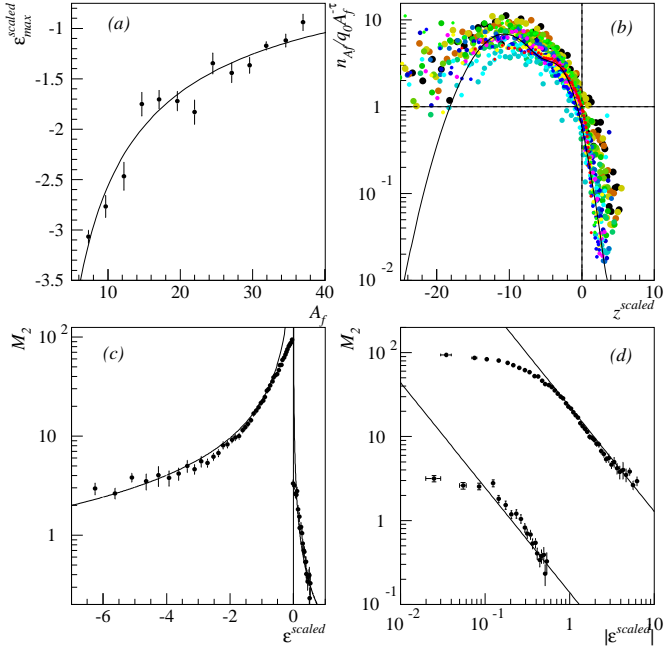,width=10.0cm,angle=0}}
\caption{Results of the system size $/$ finite size scaling analysis for Au $+$ C multifragmentation.\ \ (a) The scaled $\sigma$ 
power law.\ \ (b) The scaling function, normalized cluster distribution versus $z^{scaled}$.\ \ The solid curve show the fitted 
scaling function.\ \ (c) and (d) show the resulting $\gamma$-power law as predicted from the scaling function, $\sigma$ and $\tau$.\ \ 
Solid circles show the second moment plotted against the properly scaled ${\epsilon}^{scaled}$ and solid lines show the predicted 
$\gamma$-power law.}
\label{fig:28}
\end{figure}

Using the fitted parameterization of the scaling function and other quantities, the $\gamma$-power law can be determined as
before.\ \ Figures 28b and c show the measured $M_2$ of gold multifragmentation plotted as a function of ${\epsilon}^{scaled}$.\ \ 
A $\gamma$-power law was plotted on, not fitted, Figures 28b and c with ${\gamma} = 1.3 {\pm} 0.2$ (from $\tau$, $\sigma$ and eq.
(\ref{scaling_1})) and offsets determined via the scaled scaling function.\ \ This $\gamma$-power law agrees with the measured
$M_2$ over nearly all of the ${\epsilon}^{scaled}$-range, the exception being near ${\epsilon}^{scaled} = 0$ where finite size 
of the system limits the maximum of $M_2$.

As stated above, this analysis makes no attempt to account for the Coulomb energy of the system in the formulation of finite size 
scaling.\ \ The effect of the Coulomb energy may evidence itself in the degree of collapse of the data.\ \ The fact that, to a 
large extent, the data do collapse without any explicit adjustment to the theory ({\it i.e.} formalation of the scaling variable, 
$z^{scaled}$) may then be an indication that the Coulomb energy is not a major perturbation.\ \ Hence, it is perhaps justifiable 
to adhere to the Coulomb-free theory when scaling $T_c$.

\section{Discussion and conclusions}

The focus in the present paper has been on cluster distributions and the types of analyses which can shed light on their creation
mechanism.\ \ In particular, attempts were made to identify procedures that can distinguish those distributions which
are related to critical behavior from those which  are not.\ \ While this question is easily answered for systems containing large 
numbers of constituents, it is much more difficult to address the case of interest here, namely systems with at most only a few
hundred particles.\ \ In such systems, finite size effects play a large role.\ \ Therefore, two different computational models,
bond building percolation and a random partitions, have been used.\ \  It is well known that in the macroscopic limit, the former
system  possesses a continuous phase transition characterized by a unique scaling function and set of critical exponents while the
latter system does not.\ \ In addition, data arising from the multifragmentation of gold nuclei has been studied using the same 
procedures.\ \ For this system, it is not known, a priori, whether a critical point is present.

Many cluster properties have been proposed as being suitable measures of critical behavior.\ \ Among these are the fluctuations in
the size of largest fragment (Figure 1), peaking in the quantity ${\gamma}_2$ (Figure 3), peaking behavior in $M_2$ (Figure 5),
Campi plots (Figure 6) and simple power law behavior in the  cluster mass distribution for a particular value of the
appropriate control parameter.\ \  It was seen that none of these measures, taken alone or together, was sufficient to distinguish
a system possessing critical behavior from one which does not.

The first procedure which produced different results for critical and noncritical systems was the single parameter power law 
fit to the cluster mass distribution (Figures 12 through 15).\ \ For the percolation systems and for the multifragmentation data, 
it was shown that the one parameter power law fit describes the data well only over a very limited range of the control parameter.\ \ 
The value of the control parameter where the power law fit is best is very close to where the abovementioned peaks occur.\ \ However, 
for the random partitions, this was not the case. 

If a system possesses a critical point, it is also expected to possess a scaling function that describes its behavior away from the
critical point.\ \ Phase transition theory specifies how the argument of this function depends on cluster mass and distance from
the critical point.\ \ If such a  function exists, the theory permits the determination of the critical exponent $\sigma$.\ \ This
determination was done for the percolation system yielding satisfactory agreement with its known value.\ \ The same procedure was
applied to the random breakup model and to the multifragmentation data.\ \ It was clear from this analysis that percolation and
multifragmentation were very similar in many features, while the random system was distinctly different.\ \ See Figures 17 and 18.

Again, if a system possesses a critical point, it is expected to possess a scaling function that describes its behavior away 
from the critical point.\ \ Therefore, when the data is properly scaled, it should collapse onto a single curve.\ \ Figure 19 
shows the amount of data collapse for the systems discussed here.\ \ The quality of the data collapse (Figures 19-21) reinforces 
the notion that the random breakup system is different from the others.\ \ Although the precise form of the scaling function is 
not dictated by phase transition theory, both the percolation system and multifragmentation data were satisfactorily described by 
a sum of two gaussians.\ \ The random breakup model was not.

The issue of finite size scaling was discussed.\ \ Unlike the other systems examined here, which had a fixed number of constituents, 
the nuclear multifragmentation data originated from systems whose size varied monotonically with observed charged particle 
multiplicity.\ \ See Figure 27a.\ \ Phase transition theory makes a prediction, eq. (\ref{fss}), as to how the value of the control 
parameter at the critical point changes as a function of system size.\ \ Applying eq. (\ref{fss}) produced an improvement in the 
quality of the data collapse for the scaling function and yielded a better prediction for the behavior of $M_2$.\ \ See Figure 28.

Critical exponent values have been determined in an unbiased manner for each system.\ \ For both sets of analyses on the percolation 
clusters, the standard percolation exponents were recovered to within error bars.\ \ For the random partitioning, exponents 
could be extracted, but none that fulfilled well known scaling laws.\ \ The exponent values determined from the gold 
multifragmentation cluster distributions fulfill the scaling laws, to within error bars, and fall near the three dimensional Ising 
universality class.

The effect of secondary decays from hot initial fragments on the critical exponents has not been explicitly considered in this
paper.\ \ In the SMM \cite{bondorf_1}-\cite{botvina} such effects become significant above $E^{\ast}/A_{0} = 7 $ MeV/nucleon.\ \ 
Thus $\tau$ and $\gamma$, which are determined at lower excitation energies, will be unaffected in the SMM's fragment distributions.\ \ 
The exponent $\sigma$ is determined by the multiplicities at which individual light fragment yields attain their peak values.\ \ 
As shown in Figure 17d, the lightest fragments peak at large multiplicities, corresponding to excitation energies for which secondary 
decay are important in the SMM.\ \ An SMM calculation \cite{srivastava} indicates that the value of $\sigma$ from the SMM's fragment 
distribution was increased by about $70$\% due to this effect.\ \ However, it is unclear from that calculation that the effects of 
secondary decay are as great in the experimental data as they are in the SMM.\ \ The SMM calculation over predicts the yield of light 
fragments which could indicate that the SMM estimates of secondary decay are too severe.\ \ Corrections to the model independent 
quantities determined in this paper based on the SMM calculations are premature.\ \ See Appendix C for discussions of another set of 
model based interpretations.

Although the multifragmentation data possess many of the gross and detailed characteristics that the percolation system
does, it is not at all obvious why this should be the case.\ \ After all, real nuclei obey quantum mechanics, have varying binding
energies per particle, and, most significantly, are charged.\ \ On the other hand, it is well known that near a critical point the
details of the interaction become unimportant and only the dimensionality of the system and the dimension of the order parameter
are important.\ \ As noted in the introduction, the attractive nuclear force bears a similarity to a van der Waals force.\ \
However, the Coulomb force is a long range force and imposes a natural limit to the size of stable nuclei.\ \ Thus, it is not clear
to what extent a finite charged system can exhibit critical behavior when the macroscopic system cannot exist.\ \ The exact role of
the Coulomb force in physical systems undergoing a change of phase is currently of great interest \cite{fisher_d}, \cite{fisher_e}
and is, at this point, an open question.\ \ The philosophy of this paper has been to make use of phase transition theory as it applies 
to uncharged systems.\ \ What results for the analysis of the gold multifragmentation data bears great similarity to the results of 
the same procedures applied to a system known to possess a critical point.\ \ It is tempting then to conclude that multifragmentation 
is related to critical behavior occurring in a finite nuclear system.

\appendix
\section{Discussion of power law in random partitions}

As a demonstration of the power of the sort of scaling analysis presented above it is shown that the random partitions follow a
simple power law of $N_{A_f} \sim A_{f}^{-1}$.\ \ Figure A1 shows the scaled cluster distribution as a function of $m$ from
clusters with $A_f \ge 3$.\ \ The data nearly collapses to unity along the horizontal axis over the multiplicity range for $m >
5$.\ \ The deviations are due to the constraints of $m$ and finite size.\ \ With a simple scaling analysis the underlying power law
describing the cluster distribution becomes clear.

\begin{figure} [h]
\centerline{\psfig{file=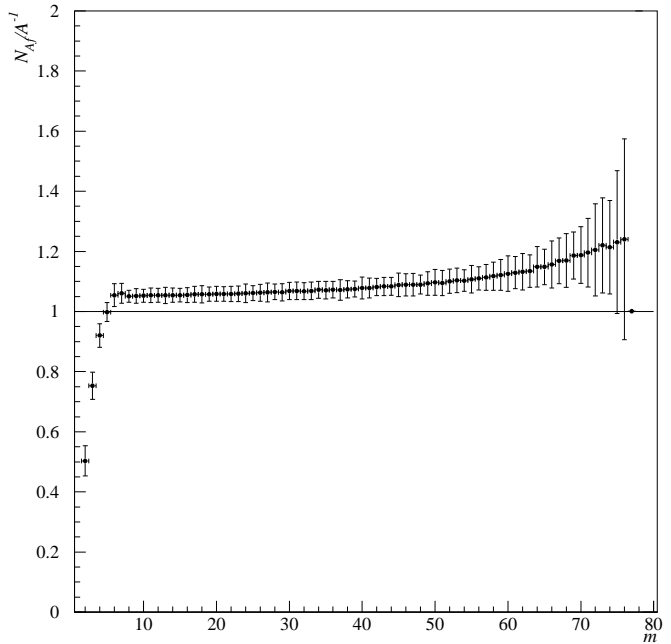,width=10.0cm,angle=0}}
\caption{Results of a scaling analysis performed on the cluster distribution of random partitions.\ \ The scaled cluster
distribution (nearly) collapses to unity over the entire multiplicity range.}
\label{fig:a1}
\end{figure}

\section{Riemann $\zeta$ function summation}

The value of $q_0$ used in this work based on the $\zeta$-function was generated with VAX FORTRAN code using double precision and
letting the sum run from $1$ to $10^6$.\ \ The sum was terminated at this point in order to keep computing times within reason.\ \
For a value of $\tau = 2.18$ summing to $10^6$ gives a value of $q_0$ that is within 10$\%$ of the value when the sum is terminated
at $10^{10}$, see Figure A2.\ \ Increasing the upper limit of the summation in the $\zeta$-function causes no significant changes
in the analysis presented in this work.

\section{Lattice gas model interpretations}

In a recent paper \cite{gc}, Gulminelli and Chomaz (GC) report on calculations made using a lattice gas model.\ \ Their results can 
be divided into two different categories: (1) an analysis of the thermodynamic quantities of the system; (2) an analysis of the 
statistical properties of the droplet or cluster distribution.

In their paper GC performed canonical calculations on a three dimensional lattice of $n$ sites ($n = 216$ or $512$) with periodic 
boundary conditions at a given temperature, $T$, and density, $\rho$.\ \ The density was varied by changing the number of particles 
in the system, $A_0$, with $\rho = A_0 / n$.\ \ For a choice of $T$, $\rho$ was varied and the system's chemical potential, $\mu$, 
was determined.\ \ At some values of $T$ the ${\mu}$ vs. ${\rho}$ (${\mu}$ vs. $A_0$) isotherm exhibited a back bend.\ \ As $T$ 
increased the back bend disappeared.\ \ This was taken as evidence that a first order phase transition (back bending ${\mu}$ vs. 
$A_0$) had culminated in a continuous phase transition (flat ${\mu}$ vs. $A_0$).\ \ The critical point, ($T_c$, ${\rho}_c$), was thus
defined.\ \ Via a Maxwell construction, GC also determined the boundary of the coexistence zone.

The theoretical ideas behind the procedure followed by GC are sound (with the exception of the questionable definition of volume in 
a system with periodic boundary conditions), however  the execution is flawed.\ \ The error in execution lies in the method by which 
$\rho$ was varied.\ \ GC varied $\rho$ by holding $n$, the volume of the system, fixed and varying $A_0$, the number of constituents 
or ``size'' of the system.\ \ Therefore, at each value of $\rho$ GC have a system of different size.\ \ It is well known that the 
effective $T_c$ of a system varies with its system's size \cite{fisher_a}-{\cite{elliott_perc_2}}.\ \ Thus each time GC varied $A_0$ 
to vary $\rho$ they were dealing with a system with a different $T_c$.\ \ With this understanding it is clear that a plot of $\mu$ 
versus $A_0$ as shown by GC in their Figure. 1a, which serves as the basis for their determination of the critical point, is difficult 
to interpret.\ \ $T_c$ cannot be deduced in the manner followed by GC.\ \ Similarly, it is impossible to determine ${\rho}_c$ for a 
system of $A_0 = 100$ using information from a system of $A_0 = 200$ without accounting for the finite size scaling of the critical 
point.\ \ This makes the thermodynamically extracted values of the critical point suspect and calls into question the validity of 
the coexistence line shown by GC in the lower part of their Figure 3.

Via an analysis of the droplet distributions GC determined the value of the critical exponent $\tau = {\tau}_{max}$ based on the peak 
in the production of different size droplets.\ \ The value of $T_c$ was determined coupling the ${\tau}_{max}$ results with a two 
parameter power law fit the droplet distribution at various temperatures, ${\tau}(T)$, at $T_c$ ${\tau}_{max} = {\tau}(T_{c})$.\ \ The 
use of two parameter fits is generally an improper method to determine $\tau$ \cite{elliott_perc_2}-{\cite{nakanishi}}.\ \ In this 
analysis GC determine $T_c$ on a system-by-system basis, avoiding the error of grouping systems of different size together.\ \ GC 
also extracted the exponent $\sigma$ and then used the above results to collapse the droplet data onto a single curve illustrating 
the scaling function that modifies the power law away from the critical point.\ \ While the data collapses onto a 
single curve, the value of the curve at $T_c$ is problematic.\ \ From GC's work the value of the scaling function at $T_c$ is 
$f(T_{c}) \sim 0.5$ which contradicts the $\tau$ values extracted by GC according to the Riemann $\zeta$-function relation between 
$\tau$ and $f(T_{c})$ \cite{nakanishi}; {\it e.g.} $\tau \sim 2.2$ gives $f(T_{c}) \sim 0.2$, while GC's $f(T_{c}) \sim 0.5$ gives 
$\tau \sim 2.7$.

Finally GC combine the results of their analyses and assert that the K\'{e}rtesz line of $\tau$ power laws continues into the 
coexistence region of the lattice gas and that this behavior is observed due to the finite size of the system and is not observed 
in systems of much greater size.\ \ However, due to the problems in GC's method of varying $\rho$ to find the coexistence curve, such 
a claim is premature.\ \ Instead, their estimates of $T_c$ from the droplet analysis, which are shown as a string of points in their 
Figure 3, may reflect the finite size scaling of the critical point and not the  K\'{e}rtesz line extending into the coexistence 
zone.\ \ That each different size system appears to be at ${\rho}_c$ as well as $T_c$, illustrated by the data collapse, may be 
explained by the relation ${\mu}-{\mu}_{coex} \sim ({\rho}-{\rho}_{c})^{\delta}$.\ \ For a 3D Ising system $\delta \sim 4.8$.\ \ 
Thus, systems of finite size near their critical density are nearly on the coexistence curve and show the sort of data collapse 
illustrated in GC's Figure 2 \cite{fisher_priv_com}.

\begin{figure} [h]
\centerline{\psfig{file=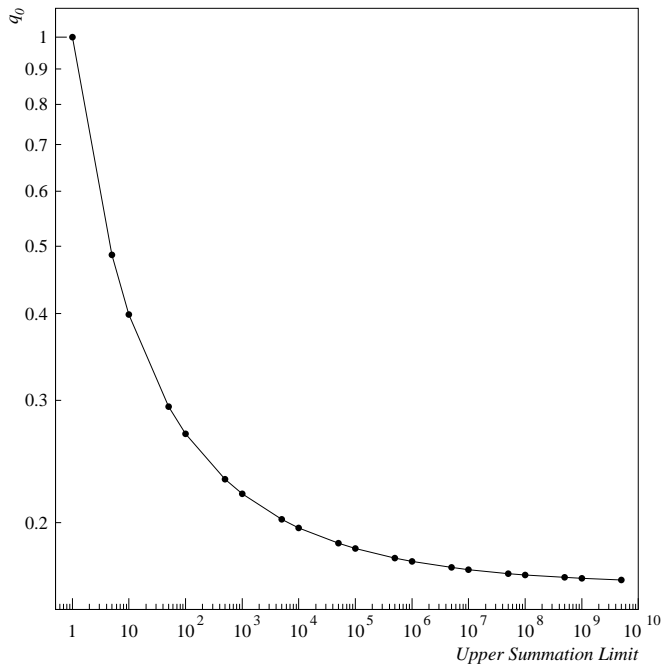,width=10.0cm,angle=0}}
\caption{Results of the summation to determine the offset $q_0$ as a function of the upper limit of the summation.}
\label{fig:a2}
\end{figure}

This work was supported in part by the U.S. Department of Energy Contracts or Grants No. DE-ACO3-76F00098, DE-FG02-89ER-40513,
DE-FG02-88ER-40408, DE-FG02-88ER40412, DE-FG05-ER40437 and by the U.S. National Science Foundation under Grant No. PHY-91-23301.


\begin{thebibliography}{}

\bibitem{finn}
	J. E. Finn {\it et al}., Phys. Rev. Lett. {\bf 49}, 1321 (1982).

\bibitem{hirsch}
	A. S. Hirsch {\it et al}., Phys. Rev. C {\bf 29}, 508 (1984).

\bibitem{fisher}
	M. E. Fisher, Physics {\bf 3}, 255 (1967).

\bibitem{fisher_2}
	M. E. Fisher, Rep. Prog. Phys. {\bf 30}, 615 (1969).

\bibitem{stauffer_kiang}
        D. Stauffer and C. S. Kiang, Advances in Colloid and Interface Science 7, 103 (1977).

\bibitem{stauffer_kiang_2}
        C. S. Kiang and D. Stauffer, Z. Physik {\bf 235}, 130 (1970).

\bibitem{minich}
        R. W. Minich {\it et al}., Phys. Lett. B {\bf 118}, 458 (1982).

\bibitem{gilkes_gamma}
	M. L. Gilkes, {\it et al}., Phys. Rev. Lett. {\bf 73}, 1590 (1994).

\bibitem{elliott_sigma}
	J. B. Elliott {\it et al}., Phys. Lett. B, {\bf 381}, 24 (1996).

\bibitem{elliott_scaling}
	J. B. Elliott {\it et al}., Phys. Lett. B. {\bf 418}, 35 (1998).

\bibitem{campi_1}
	X. Campi, J. Phys. A {\bf 19}, L917 (1986).

\bibitem{campi_2}
	X. Campi, Phys. Lett. B {\bf 208}, 351 (1988).

\bibitem{campi_3}
	X. Campi, Nucl. Phys. A {\bf 495}, 259 (1989).

\bibitem{campi_4}
	X. Campi and H. Krivine, Z. Phys. A {\bf 344}, 81 (1992).

\bibitem{campi_5}
	X. Campi and H. Krivine, Nucl. Phys. A {\bf 545}, 161 (1992).

\bibitem{campi_6}
	X. Campi and H. Krivine, Contribution to the International Workshop on Dynamical Features of Nuclei and Finite Fermi
        Systems, World Scientific Ed. (1993).

\bibitem{campi_7}
	X. Campi and H. Krivine, Contribution to the International Workshop on Multi-Particle Correlations and Nuclear Reactions,
        World Scientific Ed. (1994).

\bibitem{campi_8}
	X. Campi and H. Krivine, Nucl. Phys. A {\bf 620}, 46 (1997).

\bibitem{mekjian_1}
	H. Jaqaman, A. Z. Mekjian and L. Zamick, Phys. Rev. C {\bf 27}, 2782 (1983).

\bibitem{mekjian_2}
	H. Jaqaman, A. Z. Mekjian and L. Zamick, Phys. Rev. C {\bf 29}, 2067 (1984).

\bibitem{mekjian_3}
	A. L. Goodman, J. I Kapusta and A. Z. Mekjian, Phys. Rev. C {\bf 30}, 851 (1984).

\bibitem{mekjian_4}
	A. Z. Mekjain, Phys. Rev. Lett. {\bf 64}, 2125 (1990).

\bibitem{mekjian_5}
	A. Z. Mekjian, Phys. Rev. C {\bf 41}, 2103 (1990).

\bibitem{mekjian_6}
	S. L. Lee and A. Z. Mekjian, Phys. Rev. C {\bf 47}, 2266 (1993)

\bibitem{mekjian_7}
	K. C. Chase and A. Z. Mekjian, Phys. Rev. Lett. {\bf 75}, 4732 (1995).

\bibitem{yariv_1}
	Y. Yariv and Z. Fraenkel, Phys. Rev. C {\bf 20}, 488 (1981).

\bibitem{yariv_2}
	Y. Yariv and Z. Fraenkel, Phys. Rev. C {\bf 24}, 2227 (1989).
	
\bibitem{cugnon}
	J. Cugnon, C. Volant and S. Vuillier, Nucl. Phys. A {\bf 620}, 475 (1997).

\bibitem{toneev}
	V. D. Toneev and K. K. Gudima, Nucl. Phys. A {\bf 400}, 173 (1983).

\bibitem{danielewicz}
	P. Danielewicz, Phys. Rev. C {\bf 51}, 716 (1995).

\bibitem{peilert}
	G. Peilert {\it et al}., Rep. Prog. Phys. {\bf 57}, 533 (1994)

\bibitem{bondorf_1}
	J. P. Bondorf {\it et al}., Nucl. Phys. {\bf A}, 321 (1985)

\bibitem{bondorf_2}
	J. P. Bondorf {\it et al}., Nucl. Phys. {\bf A}, 444 (1985)

\bibitem{bondorf_3}
	J. P. Bondorf {\it et al}., Phys. Rep. {\bf 257}, 133 (1995).

\bibitem{botvina}
	A. S. Botvina {\it et al}. Nucl. Phys. A {\bf 475}, 663 (1987).

\bibitem{gross_1}
	D. H. E. Gross, Rep. Prog. Phys {\bf 53}, 605 (1990).

\bibitem{gross_2}
	D. H. E. Gross, Phys. Rep. {\bf 279}, 119 (1997).

\bibitem{schlagel}
	T. J. Schlagel and V. R. Pandharipande, Phys. Rev. C, {\bf 36}, 162 (1987).

\bibitem{aichelin}
	J. Aichelin, Phys. Rep. {\bf 202}, 233 (1991).

\bibitem{pratt}
	S. Pratt, C. Montoya and F. Ronning, Phys. Lett. B {\bf 349}, 261 (1995).

\bibitem{latora}
	V. Latora, M. Belkacem and A. Bonasera, Phys. Rev. Lett {\bf 73}, 1765 (1994).

\bibitem{donangelo}
	R. Donangelo and S. R. Souza, Phys. Rev. C {\bf 56}, 1504 (1997).

\bibitem{waddington_friar}
	C. J. Waddington and P. S. Freier, Phys. Rev. C {\bf 31}, 888 (1985).

\bibitem{gsi_1}
	C. A. Ogilvie {\it et al}., Phys. Rev. C {\bf 67}, 1214 (1991).

\bibitem{gsi_2}
	J. Hubele {\it et al}., Zeit. Phys. A {\bf 340}, 263 (1991).

\bibitem{gsi_3}
 	P. Kreutz {\it et al}., Nucl. Phys. A {\bf 556}, 672 (1993).

\bibitem{moretto}
	L. G. Moretto {\it et al}., Phys. Rep. {\bf 287}, 249 (1997).

\bibitem{phair}
	L. Phair {\it et al}., Phys. Rev. Lett. {\bf 79}, 3538 (1997).

\bibitem{theory_of_critical_phenomena}
	J. J. Binney {\it et al}., ``The Theory of Critical Phenomena'', 3rd ed. (Oxford University Press, 1995).

\bibitem{hauger_prl}
	J. A. Hauger {\it et al}., Phys. Rev. Lett. {\bf 77}, 235 (1996).

\bibitem{hauger_prc}
	J. A. Hauger {\it et al}., Phys. Rev. C {\bf 57}, 764 (1998).

\bibitem{stauffer_aharony}
	D. Stauffer and A. Aharony, ``Introduction to Percolation Theory'', 2nd ed. (Taylor and Francis, London, 1992).

\bibitem{stauffer_rep}
	D. Stauffer, Phys. Rep {\bf 54}, 2 (1979).

\bibitem{hoshen_scaling}
	J. Hoshen {\it et al}., J. Phys. A {\bf 12}, 1285 (1979).

\bibitem{leath_1}
	P. L. Leath, Phys. Rev. Lett. {\bf 36} 921 (1976)

\bibitem{leath_2}
	P. L. Leath, Phys. Rev. B {\bf 14} 5046 (1976).

\bibitem{fisher_priv_com}
	M. E. Fisher, private communication, 1996.

\bibitem{nakanishi}
	H. Nakanishi and H. E. Stanley, Phys. Rev. B {\bf 22}, 2466 (1980).

\bibitem{bonesara}
	P. F. Mastinu {\it et al}., Phys. Rev. Lett. {\bf 26}, 2646 (1996).

\bibitem{elliott_perc_1}
	J. B. Elliott {\it et al}., Phys. Rev. C {\bf 49}, 3185, (1994).

\bibitem{mastinu}
	P. F. Mastinu {\it et al}., Phys. Rev. C {\bf 57}, 831 (1998).

\bibitem{panagiotou}
	A. D. Panagiotou {\it et al}., Phys. Rev. Lett. {\bf 52}, 496 (1984).

\bibitem{bauer_1}
	W. Bauer {\it et al}., Phys. Lett. B 150, {\bf 53} (1985).

\bibitem{li_1}
	T. Li {\it et al}., Phys. Rev. Lett. {\bf 70}, 1924 (1993).

\bibitem{li_2}
	T. Li {\it et al}, Phys. Rev. C {\bf 49}, 1630 (1994).

\bibitem{gupta_1}
	S. Das Gupta and J. Pan, Phys. Rev. C {\bf 53}, 1319 (1996).

\bibitem{williams}
	C. Williams {\it et al}., Phys. Rev. C {\bf 55}, R2132 (1997).

\bibitem{margolina}
	A. Margolina {\it et al}., J. Phys. A {\bf 17}, 1683 (1984).

\bibitem{wfjm_priv}
	W. F. J. M\"{u}ller, private communication, 1996.

\bibitem{wfjm_catania}
	W. F. J. M\"{u}ller, {\it Critical Phenomena and Collective Observables}, Proceedings of the International Conference CRIS
	'96, Acicastello (Italy, 1996); Ed.s S. Costa {\it et al}., World Scientific (1996).

\bibitem{elliott_perc_2}
	J. B. Elliott {\it et al}., Phys. Rev. C {\bf 55}, 1319, (1997).

\bibitem{num_rec}
	W. H. Press, B. P. Flannery, S. A. Teukolsky, W. T. Vettering, ``Numerical Recipes'', (Cambridge University Press 1986). 

\bibitem{bauer_3}
	W. Bauer and W. A. Friedman, Phys. Rev. Lett {\bf 75}, 1297 (1995).

\bibitem{gilkes_rep_1}
	M. L. Gilkes, {\it et al}., Phys. Rev. Lett. {\bf 75}, 768 (1995).

\bibitem{guggenheim}
        E. A. Guggenheim, J. Chem. Phys., {\bf 13} 253 (1945).

\bibitem{greer}
        S. C. Greer, Phys. Rev. A, {\bf 14} 1770 (1976).

\bibitem{adler}
        J. Adler, M, Moshe and V. Privman, Annals of the Israel Physical Society, v. 5, 397 (1983).

\bibitem{saul}
	D. M. Saul, M. Wortis and D. Jasnow, Phys. Rev. B, {\bf 11} 2571 (1975).

\bibitem{guttinger}
	H. G\"{u}ttinger and D. S. Cannell, Phys. Rev. A, {\bf 24} 3188 (1981).

\bibitem{pestak}
	M. W. Pestak and M. H. W. Chan, Phys. Rev. B, {\bf 30} 274 (1984).

\bibitem{bauer_4}
	W. Bauer and A. Botvina, Phys. Rev. C {\bf 52}, R1760 (1995).

\bibitem{elliott_rep_1}
	J. B. Elliott {\it et al}., Phys. Rev. C {\bf 55}, 544 (1997).

\bibitem{fisher_a}
	M. E. Fisher and A. E. Ferdinand, Phys. Rev. Lett. {\bf 19}, 169 (1967)

\bibitem{fisher_b}
	A. E. Ferdinand and M. E. Fisher, Phys. Rev. {\bf 185}, 832 (1969)

\bibitem{fisher_c}
	M. E. Fisher and M. N. Barber, Phys. Rev. Lett. {\bf 28}, 1516 (1972).

\bibitem{barber}
	M. N. Barber, ``Phase Transitions and Critical Phenomena'' {\bf 8}, C. Domb and J. L. Lebowitz editors, 145 (Academic Press
        1983).

\bibitem{albergo}
	S. Albergo {\it et al}., Nuovo Cimento, A {\bf 89}, 1 (1985).

\bibitem{palmer_anderson}
	R. G. Palmer and P. W. Anderson, Phys. Rev. D {\bf 9}, 3281 (1974).

\bibitem{ferrenberg}
	A. M. Ferrenberg and D. P. Landau, Phys. Rev. B, {\bf 44}, 5081 (1991).

\bibitem{fisher_d}
	M. E. Fisher, J. Stat. Phys. {\bf 75}, 1 (1994).

\bibitem{fisher_e}
	Y. Levin, X. Li and M. E. Fisher, Phys. Rev. Lett. {\bf 73}, 2716 (1994).
         
\bibitem{srivastava}
	B. K. Srivastava , {\it et al}., to be published.

\bibitem{gc}
        F. Gulminelli and Ph. Chomaz, Phys. Rev. Lett. {\bf 82}, 1402, (1999).

\end{thebibliography}
\end{document}